%% file: main.tex
\documentclass[sigconf]{acmart}

\usepackage[english]{babel}
\usepackage{blindtext}
\usepackage{lipsum}
\usepackage{url} 
\usepackage{subfig}
\usepackage{multirow}
\usepackage{amssymb}

\acmYear{2019}\copyrightyear{2019}
\setcopyright{acmcopyright}
\acmConference[SIGCOMM '19]{SIGCOMM '19: 2019 Conference of the ACM Special Interest Group on Data Communication}{August 19--23, 2019}{Beijing, China}
\acmBooktitle{SIGCOMM '19: 2019 Conference of the ACM Special Interest Group on Data Communication, August 19--23, 2019, Beijing, China}
\acmPrice{15.00}
\acmDOI{10.1145/3341302.3342063}
\acmISBN{978-1-4503-5956-6/19/08}

\usepackage{xspace}

\newcommand{\jc}[1]{{\color{magenta}{\footnotesize[JC: #1]}}}

\newcommand{\camera}[1]{{\color{black} {#1}}\xspace}

\newcommand{\questions}[1]{\xspace}

\newcommand{\fillme}{{\bf XXX}~}
\newcommand{\name}{{Pano}\xspace}
\newcommand{\vr}{{360\textdegree}\xspace}
\newcommand{\vrvideo}{{\vr} video\xspace}
\newcommand{\vrvideos}{{\vr} videos\xspace}
\newcommand{\vrjnd}{{360JND}\xspace}

\newcounter{packednmbr}
\newenvironment{packedenumerate}{\begin{list}{\thepackednmbr.}{\usecounter{packednmbr}\setlength{\itemsep}{0.5pt}\addtolength{\labelwidth}{-4pt}\setlength{\leftmargin}{2ex}\setlength{\listparindent}{\parindent}\setlength{\parsep}{1pt}\setlength{\topsep}{0pt}}}{\end{list}}
\newenvironment{packeditemize}{\begin{list}{$\bullet$}{\setlength{\itemsep}{0.5pt}\addtolength{\labelwidth}{-4pt}\setlength{\leftmargin}{2ex}\setlength{\listparindent}{\parindent}\setlength{\parsep}{1pt}\setlength{\topsep}{0pt}}}{\end{list}}

\newcommand{\tightcaption}[1]{\vspace{-0.2cm}\caption{{\normalfont{\textit{{#1}}}}}\vspace{-0.2cm}}
\newcommand{\tightsection}[1]{\vspace{-0.1cm}\section{#1}\vspace{-0.0cm}}
\newcommand{\tightsubsection}[1]{\vspace{-0.2cm}\subsection{#1}\vspace{-0.0cm}}

\newcommand{\eg}{{\it e.g.,}\xspace}
\newcommand{\ie}{{\it i.e.,}\xspace}

\newcommand{\myparashort}[1]{\vspace{0.02cm}\noindent{\bf {#1}}~}
\newcommand{\mypara}[1]{\vspace{0.02cm}\noindent{\bf {#1}:}~}

\newcommand{\myparaq}[1]{\smallskip\noindent{\bf {#1}?}~}

\makeatletter
\newcommand\footnoteref[1]{\protected@xdef\@thefnmark{\ref{#1}}\@footnotemark}
\makeatother

\input{macro}

\begin{document}
\title{\name: Optimizing \vr Video Streaming with a Better Understanding of Quality Perception}



\author{\textsf{
\hspace{-1.1cm} Yu Guan$^{\star\circ}$, Chengyuan Zheng$^{\star}$, Xinggong Zhang$^{\star\circ}$, Zongming Guo$^{\star\circ}$ \hspace{1cm} Junchen Jiang\\ \vspace{0.15cm}
\hspace{2.4cm} $^{\star}$Peking University ~ $^{\circ}$PKU-UCLA JRI \hspace{3.4cm} University of Chicago
}
}









\renewcommand{\shortauthors}{Y. Guan, C. Zheng,  X. Zhang, Z. Guo, J. Jiang.}
\renewcommand{\shorttitle}{\name: Optimizing \vr Video Streaming with a Better \\Understanding of Quality Perception}

\begin{abstract}
Streaming \vrvideos requires more bandwidth than non-\vrvideos. 
This is because current solutions assume that users perceive the quality of \vrvideos in the same way they perceive the quality of non-\vr videos.
This means the bandwidth demand must be proportional to the size of the user's field of view. 
However, we found several quality-determining factors {\em unique} to \vrvideos, which can help reduce the bandwidth demand. 
They include the moving speed of a user's viewpoint (center of the user's field of view), the recent change of video luminance, and the difference in depth-of-fields of visual objects around the viewpoint. 

This paper presents {\em \name}, a \vrvideo streaming system that leverages the \vrvideo-specific factors.
We make three contributions.
(1) We build a new quality model for \vrvideos that captures the impact of the \vrvideo-specific factors. 
(2) \name proposes a variable-sized tiling scheme in order to strike a balance between the perceived quality and video encoding efficiency.
(3) \name proposes a new quality-adaptation logic that maximizes \vrvideo user-perceived quality and is readily deployable.
Our evaluation (based on user study and trace analysis) shows that compared with state-of-the-art techniques, 
\name can save 41-46\% bandwidth 
without any drop in the perceived quality, or it can raise the perceived quality (user rating) by 25\%-142\% without using more bandwidth.

\end{abstract}

\maketitle

\input{intro}

\input{motivate}

\input{overview}

\input{jnd}

\input{tiling}

\input{control}

\input{impl}

\input{eval}

\input{limitations}

\input{related}



\tightsection{Conclusion}
High-quality \vrvideo streaming can be prohibitively bandwidth-consuming. 
Prior solutions have largely assumed the same quality perception model as traditional non-\vr videos, limiting the room for improving \vrvideos by the same bandwidth-quality tradeoffs as traditional videos.
In contrast, we show that users perceive \vrvideo quality differently than that of non-\vr videos. 
This difference leads us to revisit several key concepts in video streaming, including perceived quality metrics, video encoding schemes, and quality adaptation logic.
We developed \name, a concrete design inspired by these ideas.
Our experiments show that \name significantly improves the quality of \vrvideo streaming over the state-of-the-art, \eg 25\%-142\% higher mean opinion score with same bandwidth consumption).

\section*{Acknowledgment}
We thank our shepherd Zafar Ayyub Qazi and SIGCOMM reviewers for their helpful feedback.
Xinggong Zhang was supported in part by Project 2018YFB0803702, ALI AIR project XT622018001708, and iQIYI. Junchen Jiang was supported in part by a Google Faculty Research Award and CNS-1901466. Xinggong Zhang is the corresponding author.



\vspace{-0.2cm}

\appendix
\input{appendix}

\end{document}

%% file: macro.tex
\newcommand{\Tile}{\ensuremath{t}\xspace}
\newcommand{\Quality}{\ensuremath{q}\xspace}
\newcommand{\PSPNR}{\ensuremath{P}\xspace}
\newcommand{\PMSE}{\ensuremath{M}\xspace}
\newcommand{\JND}{\ensuremath{JND}\xspace}
\newcommand{\Base}{\ensuremath{C}\xspace}
\newcommand{\Action}{\ensuremath{A}\xspace}
\newcommand{\Efficiency}{\ensuremath{\gamma}\xspace}
\newcommand{\Chunk}{\ensuremath{k}\xspace}
\newcommand{\Bitrate}{\ensuremath{R}\xspace}
\newcommand{\bitrate}{\ensuremath{r}\xspace}
\newcommand{\Area}{\ensuremath{S}\xspace}
\newcommand{\Pixel}{\ensuremath{p}\xspace}

%% file: intro.tex

\tightsection{Introduction}


\vr videos are coming to age, with most major content providers offering \vr video-based applications~\cite{facebook,Google,Hulu,Vimeo,Netflix,Youku,iQiyi}.
At the same time, streaming \vrvideos is more challenging than streaming traditional non-\vrvideos.
To create an immersive experience, a \vrvideo must stream the content of a large sphere, in high resolution and without any buffering stall~\cite{att,Gaddam}.
To put it into perspective, let us consider a traditional full-HD video of 40 pixels per degree (PPD) displayed on a desktop screen, which is an area of $\sim$48{\textdegree} in width 
as perceived by viewer's eyes (if the screen is 15" in width at a distance of 30" to the viewer). 
Streaming this video on the laptop screen takes roughly 5~{Mbps}. 
In contrast, if we want to keep the perceived quality level (same PPD) for the panoramic sphere, it will take 400~{Mbps}, 80$\times$ more bandwidth consumption~\cite{VRBitrate}.

This paper is motivated by a simple, yet seemingly impossible quest:
can we stream \vrvideos in the same perceived quality as traditional non-\vr videos {\em without} using more bandwidth?
Given that today's Internet is capable of streaming high-quality videos to billions of users in most parts of the world, achieving this goal would have great societal implications and could spur massive popularization of \vr videos.

Unfortunately, the current approaches fall short of achieving this goal. 
Most solutions (\eg~\cite{Flare, ClusTile, viewpoint-driven1, viewpoint-driven5, Fan2017Fixation, Corbillon2017Viewport, Sreedhar2017Viewport}) follow the viewport-driven approach, where only the {\em viewport} (the region facing the viewer) is streamed in high quality, but this approach has several limitations.
First, a viewport ($\sim$110{\textdegree} in width~\cite{Oculus}) is still much larger than a laptop screen ($\sim$48{\textdegree} in width) as perceived by users,
so to stream a viewport region would still need at least twice the bandwidth of streaming a screen-size video at the same quality~\cite{Res_Rate}.
Second, as the viewport content needs to be pre-fetched, the player must predict where the user will look at in the future, so any prediction error can cause playback rebuffering or quality drops.
Third, to adapt to arbitrary viewport movements, the \vrvideo must be spatially split into small tiles, which could substantially increase the size of the video.

In this work, we look beyond the viewport-driven approach 
and show that the quality of \vrvideos is perceived {\em differently} than that of non-\vr videos, due to the presence of viewpoint movements\footnote{This paper makes two assumptions: 
(1) the movement of the head-mounted device can approximate the movement of the actual viewpoint, and 
(2) the object closest to the viewpoint is the one being watched by the user. 
These assumptions might be simplistic, but they can be refined with recent work on accurate viewpoint tracking (\eg~\cite{tobii, smarteye, yu2015framework, lo2017360, corbillon2017360}).}.
In particular, we empirically show three quality-determining factors {\em unique} to \vrvideos.
The user's sensitivity to the quality of a region $M$ is dependent on 
(1) the {\em relative viewpoint-moving speed} between the movement of viewpoint (center of the viewport) and the movement of visual objects in the region $M$,
(2) the {\em difference of depth-of-field (DoF)} between the region $M$ and the viewpoint-focused content, 
and (3) the {\em change in luminance} of the viewport in the last few seconds.
For instance, when the viewpoint moves slowly (\eg $<$5 deg/s), users tend to be sensitive to small quality distortion; when the viewpoint moves quickly (\eg shaking head or browsing landscape), the sensitivity can drop sharply---users might be insensitive to large quality distortion. 
In short, how sensitive a user is to quality distortion can {\em vary} over time due to the viewpoint movements.
(See \S\ref{subsec:opportunities} for more discussions.)

The observation that users perceive \vr video quality differently opens up
new opportunities to improve \vrvideo quality and save bandwidth. 
If we know a user's sensitivity to quality distortion, we can raise quality by a maximally perceivable amount, when there is spare bandwidth; and we can lower the quality by a maximal yet imperceptible amount, when the bandwidth is constrained. 
The underlying insight is that each user has a {\em limited span of attention}.
For instance, when a user moves her viewpoint, the area being watched does increase, but since the attention will be spread across a wider area, the user's attention {\em per-pixel} actually decreases. 


To explore these opportunities, this paper presents {\em \name}, a \vr video streaming system that entails three contributions:

{\em First, \name is built on a new quality model for \vrvideos that systematically incorporates the new quality-determining factors (\S\ref{sec:jnd})}.
We run a user study\footnote{Our study was IRB approved by our university, IRB00001052-18098. It does not raise any ethical issues.} to quantitatively show the relationship between the user's sensitivity to quality distortion and the relative viewpoint-moving speed, the difference of depth-of-field (DoF), and the change of luminance.
The new model allows us to estimate the subjectively perceived video quality more accurately than traditional video quality metrics (\eg PSNR~\cite{PSNR}).

{\em Second, \name uses a novel variable-sized tiling scheme to cope with the heterogeneous distribution of users' sensitivity over the panoramic sphere (\S\ref{sec:tiling}).}
Traditionally, a \vrvideo is split into equal-sized tiles (\eg 6$\times$12, 12$\times$24), each encoded in multiple quality levels, so that the player can choose different quality levels for different tiles as the viewport location moves.
This uniform tiling scheme, however, 
might be either too coarse-grained to reflect where the user sensitivity varies, or too fine-grained to contain the video encoding overhead.
Instead, \name uses {\em variable}-sized tiling scheme, which splits the video into tiles of different sizes so that a user tends to have similar sensitivity when watching the same tile.

\begin{figure}[t]
  \centering
  \includegraphics[width=0.3\textwidth]{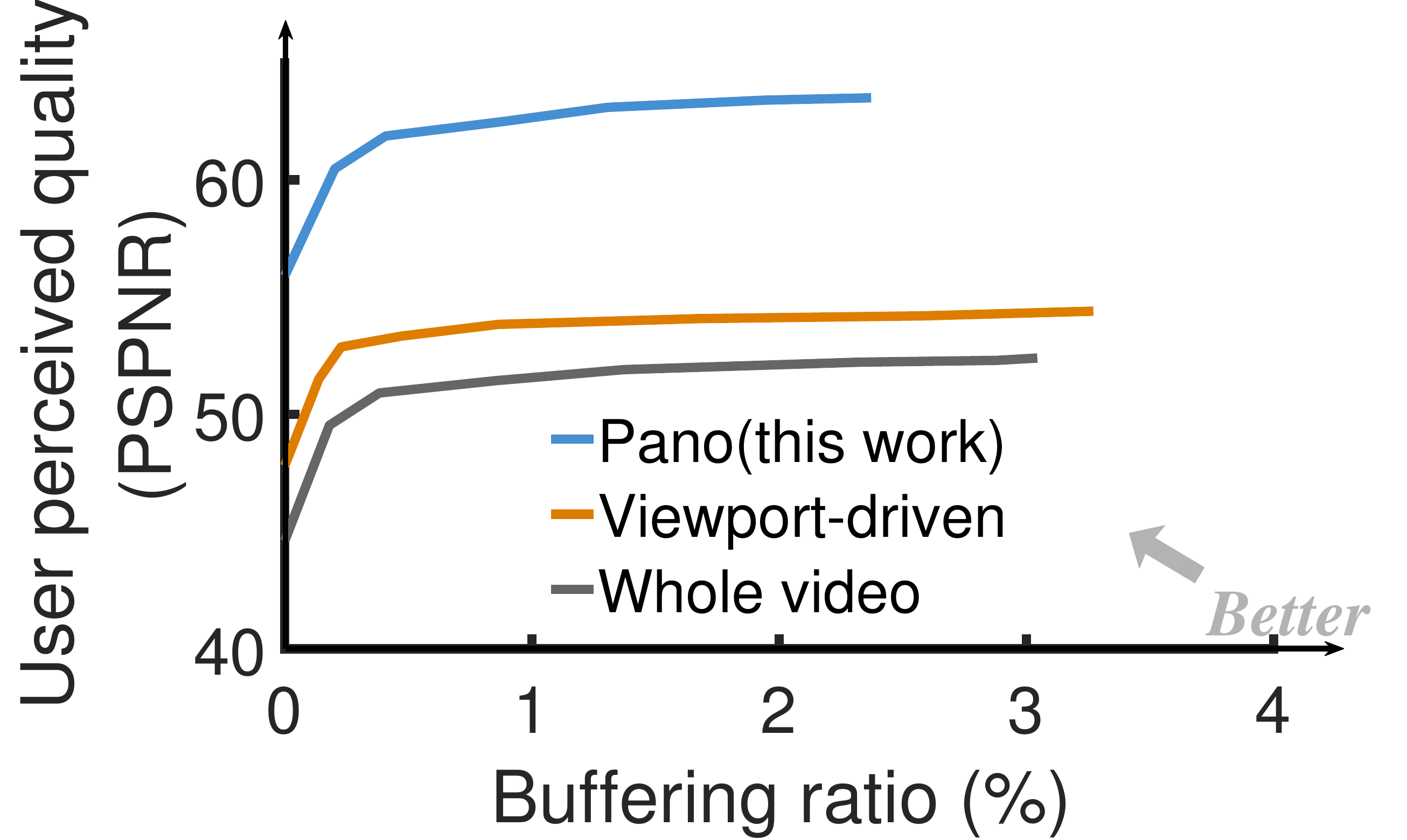}
  \tightcaption{Performance of \name and the popular viewport-driven approach on 18 \vr videos with real viewpoint traces over an emulated cellular network link.
  Full results are in \S\ref{sec:eval}. 
  \vspace{-0.3cm}
  }
  \label{fig:intro}
\end{figure}

{\em Finally, \name adapts video quality in a way that is (a) robust to the vagaries of viewpoint movements, and (b) readily deployable in the existing video delivery infrastructure (\S\ref{sec:control}).}
\name optimizes user-perceived quality by dynamically predicting viewpoint movements and adapting quality accordingly.
Despite the inevitable viewpoint prediction errors, \name can still pick the desirable quality levels, because to estimate the user's sensitivity to quality distortion, it suffices to predict the {\em range} of viewpoint-moving speed, luminance and DoF.
In addition, since \name needs information from both client (\ie viewpoint trajectory) and server (\ie video pixel information), it is incompatible with the mainstream DASH protocols~\cite{DASH} where a client locally makes bitrate-adaptation decisions. 
To address this, \name~{\em decouples} the bitrate adaptation into an offline phase and an online phase.
The offline phase pre-computes the perceived quality estimates under a few carefully picked viewpoint movements, and then it sends them to the client at the beginning of a video. 
In the online phase, the client predicts the perceived quality by finding a similar viewpoint movement that has a pre-computed estimate.

We implemented a prototype of \name 
and evaluated it using a combination of user studies (20 participants, 7 videos) and trace-driven simulations (48 users, 18 videos).
Across several content genres (\eg sports, documentary), \name can increase the mean opinion score (MOS)~\cite{MOS} by 25-142\% over a state-of-the-art solution without using more bandwidth. 
It can also save bandwidth usage by up to 46\% or reduce buffering by 60-98\% without any drop in perceived quality. 
\name suggests a promising {\em alternative} to the popular viewport-driven approach (\eg Figure~\ref{fig:intro}), which could potentially close the gap of bandwidth consumption between \vr videos and traditional videos as we have hoped.



%% file: motivate.tex

\tightsection{Motivation}
\label{sec:motivate}

\begin{figure*}[t!]
  \centering
  \includegraphics[width=0.92\textwidth]{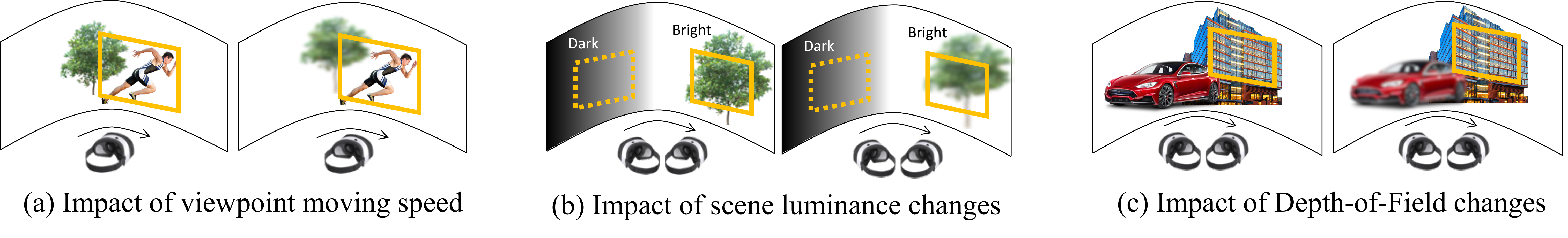}
  \vspace{-2mm}
  \tightcaption{Illustrative examples of three \vrvideo quality-determining factors, and how they help save bandwidth by reducing the quality of some part of the video without affecting the user-perceived quality. The yellow boxes indicate the viewport area (dashed ones are the previous viewport). In each case, the left-hand side and the right-hand side have similar perceived QoE, despite quality distortion on the right. 
  }
  \label{fig:examples}
  \end{figure*}

We begin by setting up the background of \vrvideo streaming (\S\ref{subsec:background}).
Then we introduce the quality-determining factors unique to \vrvideos (\S\ref{subsec:opportunities}), and analyze the potential improvement (\S\ref{subsec:potentials}) of leveraging these factors.

\tightsubsection{Background of \vr video streaming}
\label{subsec:background}

There are already 36.9 million VR users in the US (over 10\% of its population)~\cite{eMarketer}.
By 2022, there will be 55 million active VR headsets in the US, as many as Netflix members in the US in 2018~\cite{qz}.
Many content providers (YouTube~\cite{Google}, Facebook~\cite{facebook}, Netflix~\cite{Netflix}, Vimeo~\cite{Vimeo}, Hulu~\cite{Hulu}, iQIYI~\cite{iQiyi}) offer \vrvideo streaming services on various platforms~\cite{Oculus, samsung, daydream}.

\camera{The proliferation of \vrvideos is facilitated in part by the cheap and scalable delivery architecture.\questions{B.8}}
Like other Internet videos, \vrvideos can be delivered to viewers through content delivery networks (CDNs).
A \vrvideo is first converted to a planar video and encoded by a \vr encoder (\eg~\cite{ffmpeg}), which transcodes and chops the video into chunks (or segments); 
these video chunks are then sent to geo-distributed CDN servers; and finally, a client (VR headset or smartphone) streams the video chunks sequentially from a nearby CDN server using the standard HTTP(S) protocols~\cite{DASH,hls,WOWZA}.
To cope with bandwidth fluctuations, each video segment is encoded in different quality levels, such as quantization parameters (QP), so that during playback the player can dynamically switch between quality levels at the boundary of two consecutive chunks, similar to traditional bitrate-adaptive streaming.

A distinctive feature of \vr video streaming is that the viewer's attention is {\em unevenly} distributed, with more attention in the {\em viewport} area (which directly faces the user) than the rest of the video. 
In contrast, non-\vr videos are displayed in a more confined area (\eg a desktop screen), so the uneven distribution of attention is less obvious.
The uneven distribution of attention has spurred a rich literature around the idea of {\em viewport-driven} streaming (\eg~\cite{Gaddam,Graf,POI360,Flare}) to improve \vrvideo quality. 
It spatially partitions a video into {\em tiles} (\eg 6-by-12 grids) and encodes each tile in multiple quality levels, so the \vrvideo player can dynamically assign a higher quality level to tiles closer to the viewpoint (the center of a viewport). 
Unfortunately, viewport-driven streaming has two limitations.
\camera{First, like traditional videos, each \vrvideo chunk must be prefetched before the user watches it, but viewport-driven streaming only fetches the viewport region in the hope that the fetched content matches the user's viewport. 
So any viewpoint prediction error may negatively affect user experience.\questions{B.1}}
Second, to assign quality by the distance to the dynamic viewpoint, the video must be split into many fine-grained tiles~\cite{Flare}
or encoded in multiple versions each customized for certain viewpoint trajectory~\cite{ClusTile}, but both methods could significantly increase the video size.
\vspace{-0.08cm} 
\tightsubsection{New quality-determining factors}
\label{subsec:opportunities}

A basic assumption underlying the prior efforts is that users perceive the quality of \vrvideos (within the viewport) in the same way they perceive the quality of non-\vrvideos. 
This assumption limits the room for improving the performance of streaming \vrvideos.
In other words, since the viewport appears larger than a desktop screen to the user, it still takes more bandwidth to stream a \vrvideo than a traditional screen-size video.

In contrast, our key insight is that the user-perceived quality of \vrvideos is uniquely affected by users' viewpoint movements. 
Here, we explain three quality-determining factors that are induced by a user's viewpoint movements (readers may refer to \S\ref{sec:jnd} for more analysis of their impacts on quality perception). 


\begin{packeditemize}

\item {\bf Factor~\#1: Relative viewpoint-moving speed.} 
In general, the faster the user's viewpoint moves, the less sensitive the user is to quality distortion.
Figure~\ref{fig:examples}(a) illustrates how this observation could help save bandwidth: when the user moves her viewpoint, reducing the quality level of the static background will have little impact on user-perceived quality.
Of course, the moving objects being tracked by the viewpoint will now appear static to the user, so its quality degradation has a negative impact on the perceived quality.
\camera{This idea is particularly relevant to sports videos, where the viewpoint often moves with fast-moving objects.}\questions{A.9}


\item {\bf Factor~\#2: Change in scene luminance.}
As a user moves her view around, the viewed region may switch between different levels of luminance; when the content changes from dark to bright (and vice versa), users tend to be less sensitive to quality distortion in a short period of time (typically 5 seconds~\cite{darkadaptation, darkadaptation2}).
Figure~\ref{fig:examples}(b) illustrates a simple example of how one can carefully lower the quality level of part of the video without causing any drop in the user-perceived quality.
\camera{Luminance changes are prevalent in urban night scenes, where the viewpoint may switch frequently between different levels of brightness.}\questions{A.9}


\item {\bf Factor~\#3: Difference in depth-of-field (DoF).} 
In \vrvideos, users are more sensitive to quality distortion of a region whose DoF\footnote{\vr displays can simulate DoF by projecting an object to two eyes with a specific binocular parallax (disparity)~\cite{DoF1,DoF2}.} 
is closer to that of the viewpoint.
So, users may have different sensitivities to the quality of the same region, depending on the DoF of the current viewpoint.
As illustrated in Figure~\ref{fig:examples}(c), one can save bandwidth by dynamically tracking the DoF of the viewpoint and reducing the quality level of objects that have great difference in DoFs to the viewpoint.
\camera{DoF adaptation tends to benefit outdoor videos where the viewpoint may switch between foreground objects (low DoF) and scenic views (high DoF).}\questions{A.9}



\end{packeditemize}






\vspace{0.1cm}
\mypara{Intuitive explanation}
The key to understanding these opportunities is that each user has {\em a limited span of attention}.
Although the video size grows dramatically to create an immersive experience, a user's span of attention remains largely constant.
As a result, a user often gives less attention to the specifics of a \vr video, which in turn reduces her sensitivity to quality distortion.

\myparaq{What is new about them}
\camera{Although prior work (\eg~\cite{encodingDoF1,encodingDoF2,hadizadeh2013saliency, lee2012perceptual}) also improves video encoding and streaming by leveraging the video perceptual features (\eg luminance and salient objects) and intrinsic dynamics (\eg fast changing content),
it is always assumed that 
these factors are determined by the video content, {\em not} users' viewpoint movements. 
In contrast, we seek to take into account the object movements, luminance changes, and DoF differences, all caused by users' viewpoint movements, so our approach can be viewed complementary to this body of prior work. \questions{A.1}}
For instance, static objects may appear as fast moving objects to a \vrvideo user (thus can tolerate low quality), if the user moves the viewport rapidly.
Similarly, fast moving objects will appear static to the user (thus requiring high quality), if her viewpoint moves with the object.

\tightsubsection{Potential gains}
\label{subsec:potentials}

Next, we use real viewpoint traces to demonstrate the potential benefits of these quality-determining factors.
The traces~\cite{VRdataset_page} consist of 864 distinct viewpoint trajectories (18 \vrvideos each watched by 48 users~\cite{vive}, see Table~\ref{tab:dataset} for a summary).
We measure viewpoint-moving speed in degrees per second (deg/s), luminance in gray level~\cite{luminance1,PSPNR}, and DoF in dioptres~\cite{DoF1,DoF2}.

\begin{figure}
 \setlength{\belowcaptionskip}{-0cm}
  \centering
   \begin{minipage}[t]{0.33\linewidth}
\centering
\includegraphics[width=\linewidth]{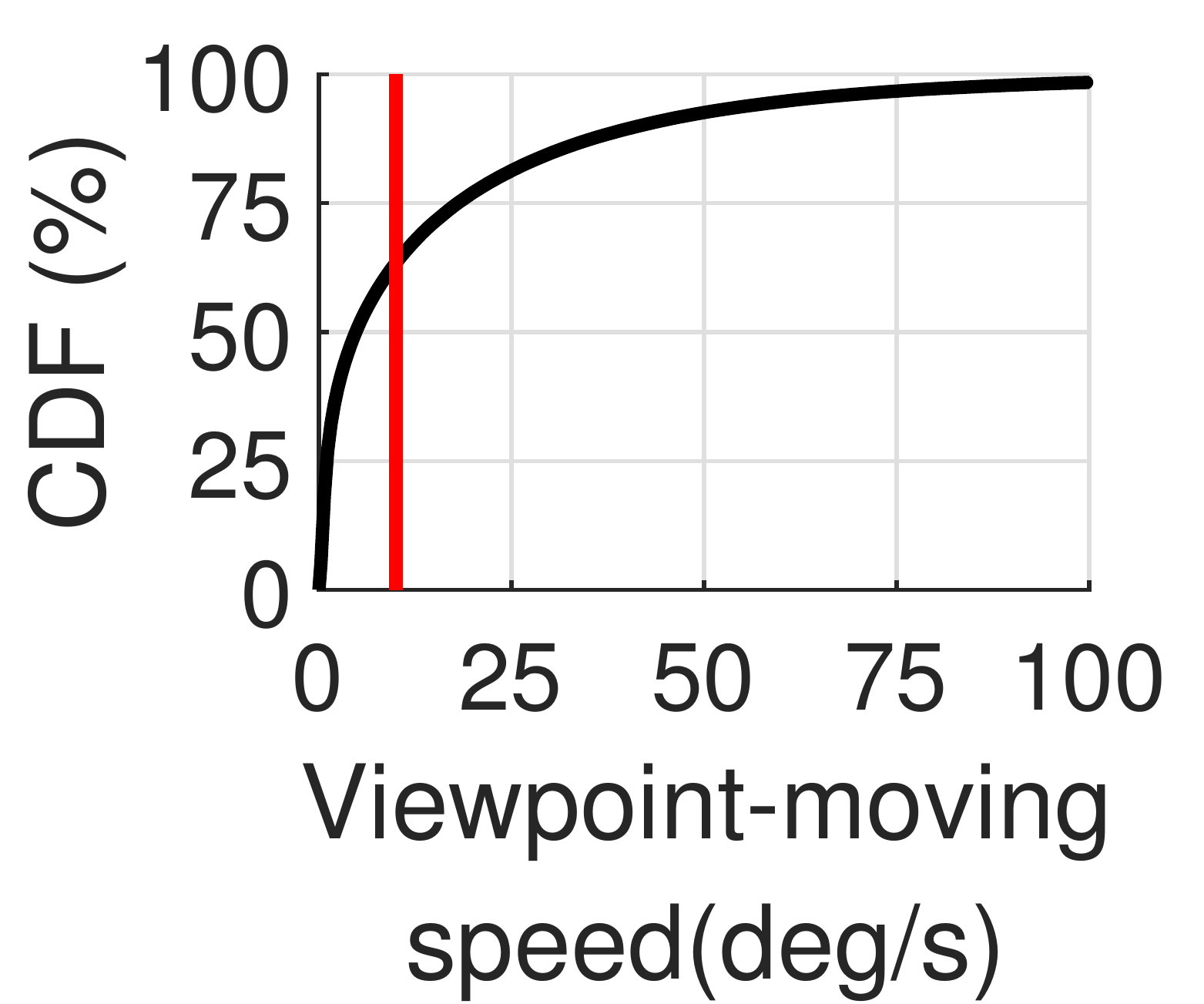}
\label{fig:side:a}
\end{minipage}
\hspace{-0.2cm}
\begin{minipage}[t]{0.33\linewidth}
\centering
\includegraphics[width=\linewidth]{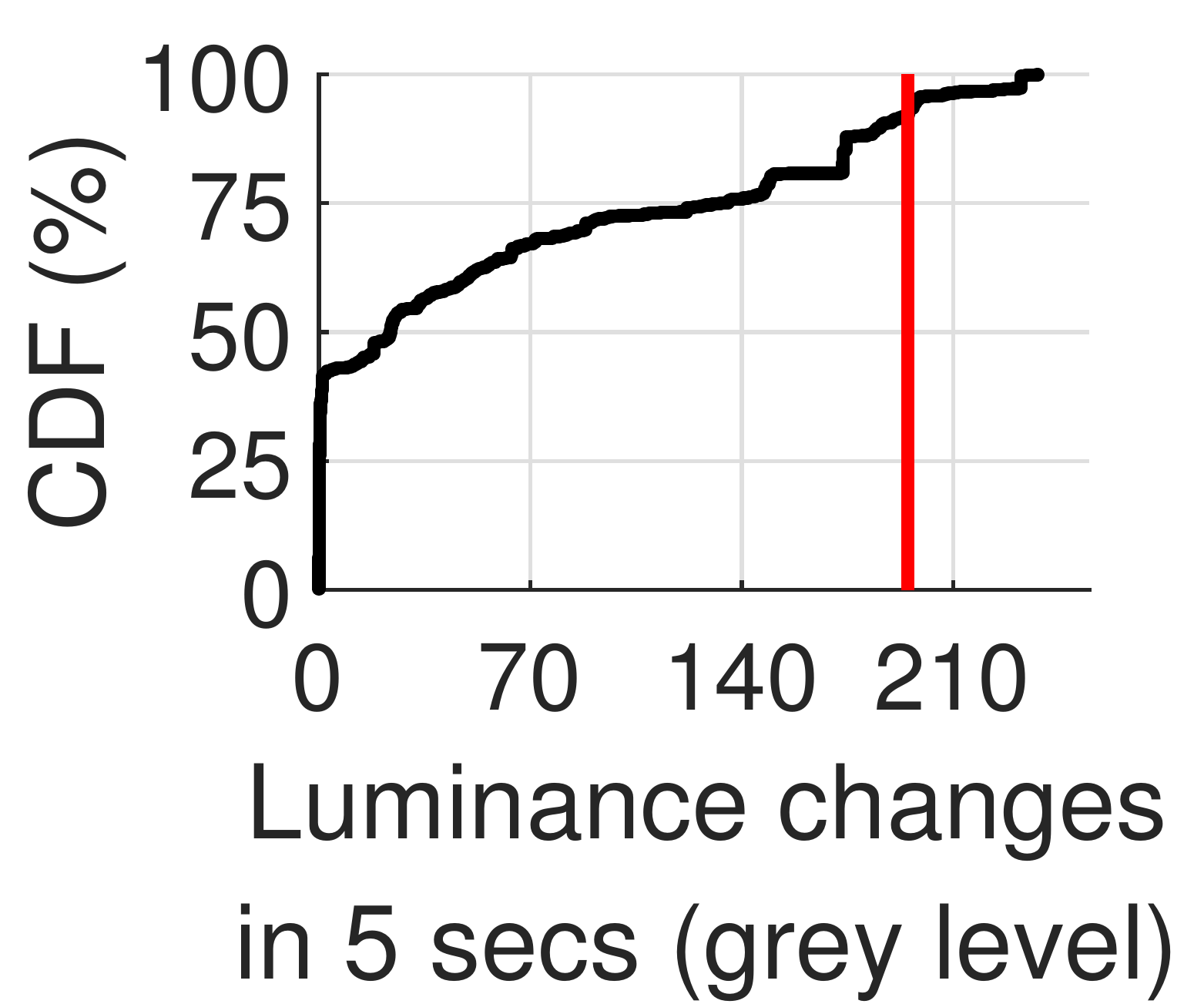}
\label{fig:side:b}
\end{minipage}
\hspace{-0.2cm}
\begin{minipage}[t]{0.34\linewidth}
\centering
\includegraphics[width=\linewidth]{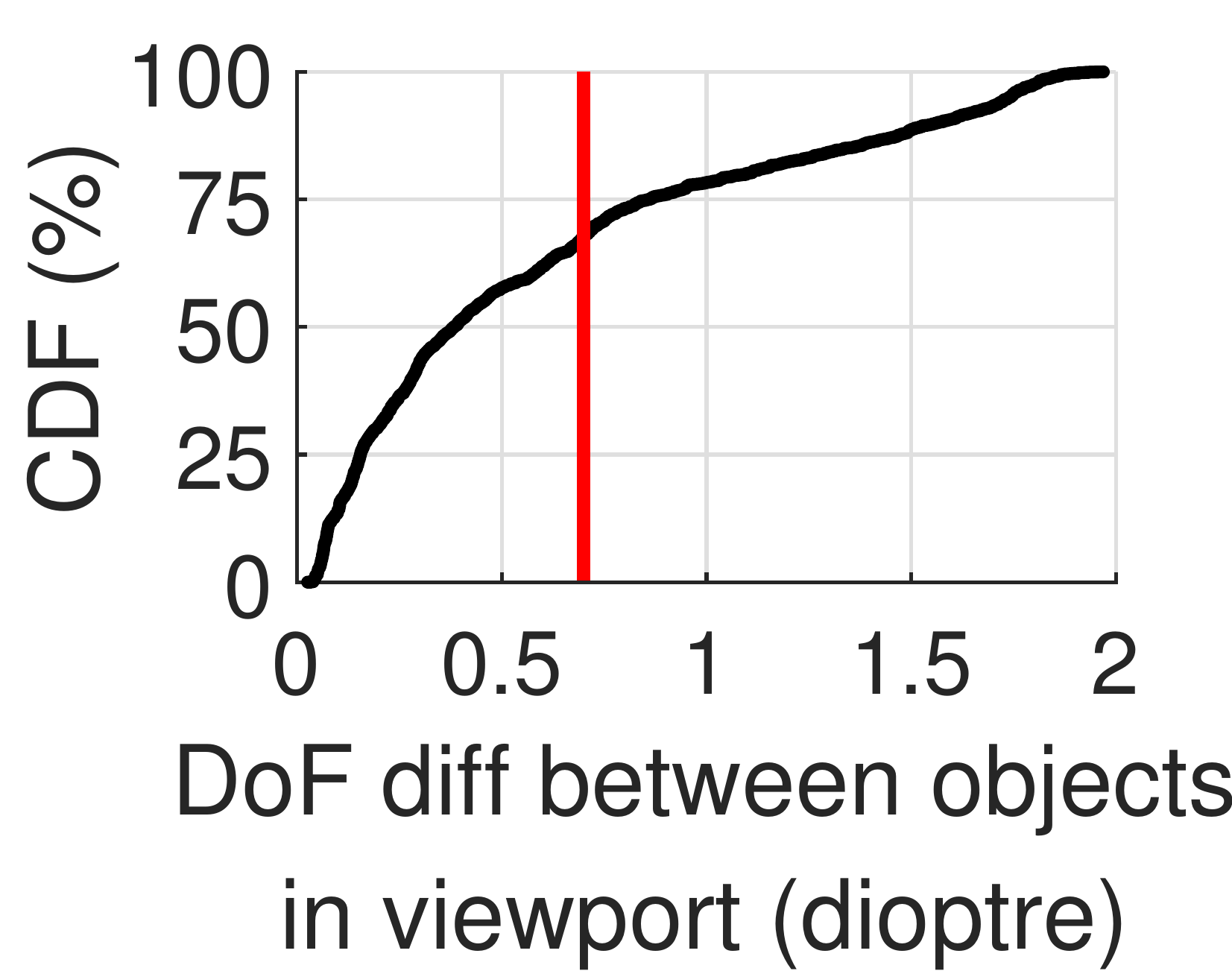}
\label{fig:side:c}
\end{minipage}
\vspace{-0.3cm}
  \tightcaption{Distribution of the new quality-determining factor values.}
  \vspace{-0.2cm}
  \label{fig:prevalence}
 \end{figure}

Figure~\ref{fig:prevalence} shows the distribution of viewpoint-moving speeds, the distribution of maximum luminance changes in different 5-second time windows, and the distribution of maximum DoF differences between two regions in one frame. 
To see how these values impact users' sensitivities to quality distortion, we measure how often these values exceed some thresholds so that users can tolerate 50\% more quality distortion than they would have if the viewpoint was static.
Based on our empirical user study in \S\ref{subsec:jnd:details}, such threshold of viewpoint-moving speed is 10 deg/s, that of luminance change is 200 gray level, and that of DoF difference is 0.7 diopters.

We can see that all three factors exceed their thresholds for 5-40\% of time.
In other words, for instance, for 40\% of time the viewpoint moves over 10 deg/sec, which means during that time, the users can tolerate 50\% more quality distortion on background pixels than they would have if the video is viewed on a computer screen.
\camera{It should be noticed that the viewpoint movements appear to be dynamic, in part because the dataset includes many outdoor sports and adventure videos.
\questions{B.4}}

%% file: overview.tex

\tightsection{\name overview}

Exploring the aforementioned opportunities, however, requires not only changing the objective of video quality optimization, but also {\em re-architecting} several critical components of the video streaming system. 
We present {\em \name}, a \vrvideo streaming system that addresses three key challenges.

\vspace{0.2cm}
\myparashort{\em Challenge 1: How to predict \vrvideo user-perceived quality by incorporating these new quality-determining factors?}
To our best knowledge, none of the existing video quality metrics directly captures the three new factors, so we first need to augment the existing video quality metrics to measure different user sensitivities under different viewpoint trajectories. 

\noindent{\bf Our solution:} 
\name presents a novel \vrvideo quality metric (\S\ref{sec:jnd}) that models the users' sensitivities to quality distortion as a function of viewpoint-moving speed, luminance change, and DoF difference.
A naive approach would profile all possible combinations of these values and each video. 
Fortunately, we show that we can {\em decouple} the impact of these factors driven by viewpoint movements from the impact of the video content. 
Moreover, we found that the impact of individual factors is largely mutually independent, which further reduces the efforts to build the new \vrvideo quality metric.

\begin{figure}
\centering
\includegraphics[width=0.34\textwidth]{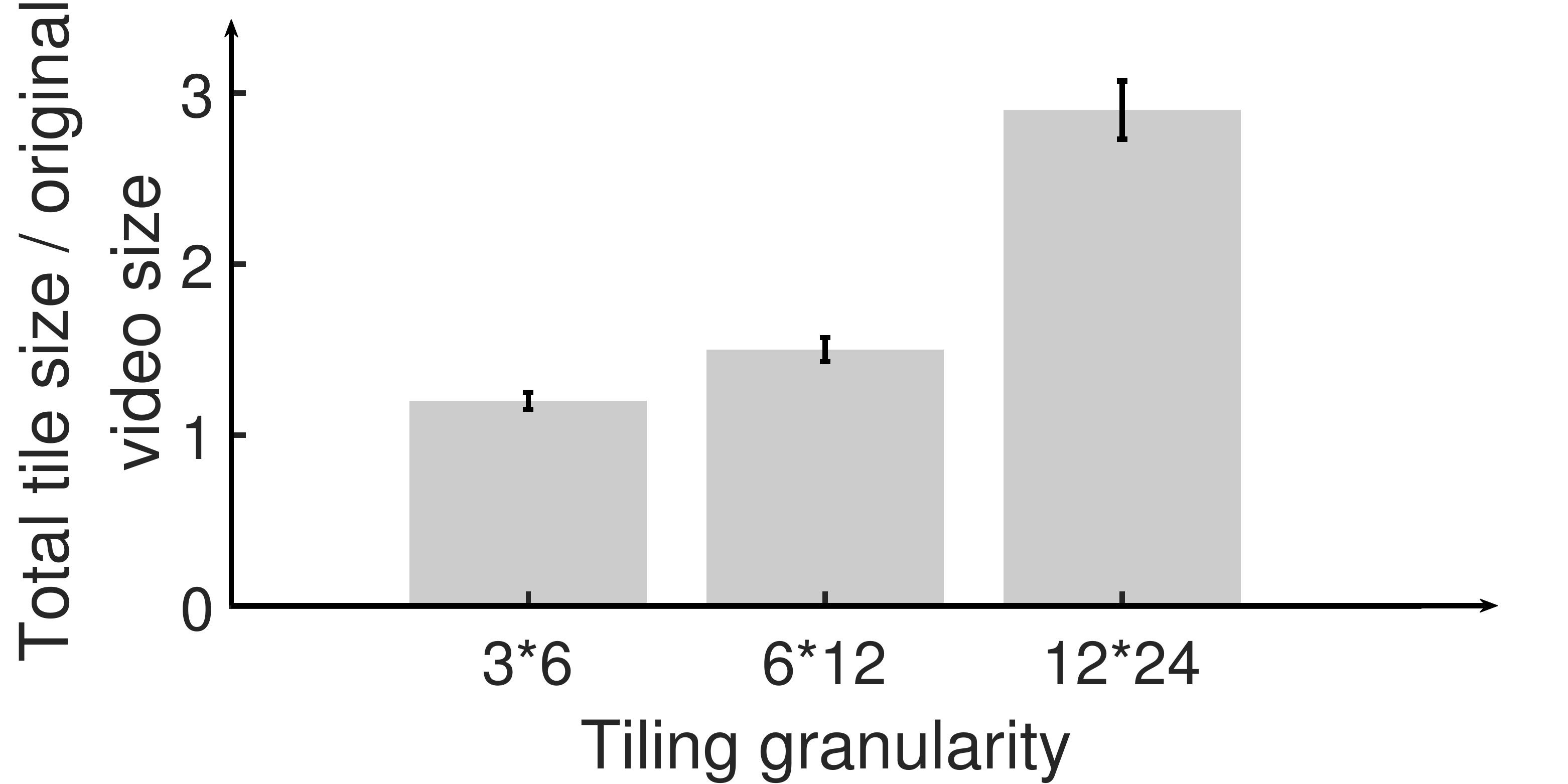}
\tightcaption{Average video sizes under different tiling granularities. (Error bars show the standard deviation of mean). 
\vspace{-0.4cm}
}
\label{fig:bitrate-efficiency}
\end{figure}

\vspace{0.2cm}
\myparashort{\em Challenge 2: How should the \vrvideos be spatially split into tiles to better exploit the new opportunities?}
Ideally, the tiling should separate regions with different object-moving speeds (\eg foreground moving objects vs. static background), different DoF, or different luminance values.
But naively splitting the video into small tiles (\eg 12$\times$24) will increase the video size by almost 200\% compared to a coarser 3$\times$6-grid tiling (Figure~\ref{fig:bitrate-efficiency}).

\noindent{\bf Our solution:} 
\name splits it into a handful of {\em variable-size} tiles (\S\ref{sec:tiling}), rather than equally sized tiles  (see Figure~\ref{fig:tiling-steps} for an example).
As a result, users tend to have similar sensitivities to quality distortion within each tile (according to history trajectories traces). 
In this way, we can maintain a coarse tiling granularity to save bandwidth while still being able to assign higher quality where users are more sensitive.

\vspace{0.2cm}
\myparashort{\em Challenge 3: How to adapt quality in a way that is robust to dynamic viewport movements and readily deployable over the existing delivery infrastructure?}
The video quality adaptation strategy needs to be revisited for two reasons.
First, it must tolerate the vagaries of available bandwidth {\em and} the inevitable errors of viewpoint movement prediction. 
Second, it must be deployable on the existing client-driven video streaming protocol~\cite{DASH}, but if done naively,  one would need {\em both} client-side information (current viewpoint movement) and server-side information (video content) to determine the sensitivity of a user to quality distortion.

\noindent {\bf Our solution:} 
Our empirical study shows that to pick the desirable quality for each tile, it is sufficient to estimate a {\em range} of the viewpoint movement, rather than their precise values (\S\ref{subsec:adaptation}).
For example, if the viewpoint moves quickly in a short time, it will be difficult to predict the exact relative viewpoint-moving speed, but \name can still reliably estimate a {\em lower bound} of the speed based on recent history.
Although \name may lose some performance gains (\eg assigning a higher-than-necessary quality given underestimated relative viewpoint-moving speeds), the conservative decisions still outperform the baselines which ignore the impact of viewpoint movements.
Finally, to be compatible with the existing client-driven video streaming architecture, \name encodes a look-up table in the video manifest file so that the client can approximately estimate the perceived quality of each quality level without accessing the actual video content (\S\ref{subsec:compatible}).

\begin{figure}[t!]
\centering
\includegraphics[width=0.47\textwidth]{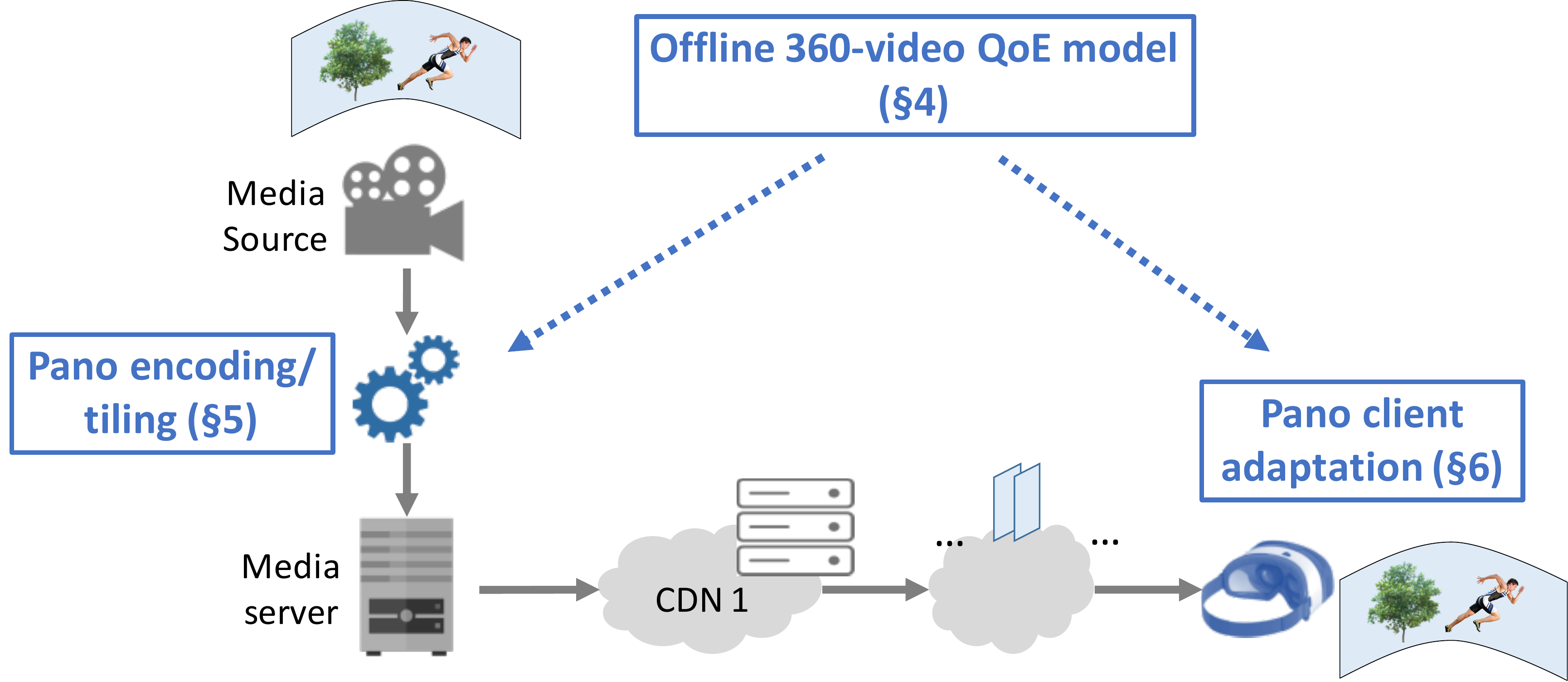}
\tightcaption{Overview of \name and how it fits in the \vr video delivery. \vspace{-0.1cm}}
\label{fig:overview}
\end{figure}

As shown in Figure~\ref{fig:overview}, although a video delivery system involves many comments, deploying \name only requires minor changes by the content provider (who controls the video encoding) and client-side device (usually also managed by the same content provider). No change is needed to the CDNs or the HTTP streaming protocol.

%% file: jnd.tex
\tightsection{\name: \vrvideo quality model}
\label{sec:jnd}

We start with \name's video quality model, which estimates the user-perceived quality under certain viewpoint movement.

\tightsubsection{A general video quality framework}
\label{subsec:jnd:framework}


%

Conceptually, \name incorporates the new quality-determining factors in Peak Signal-to-Perceptible-Noise Ratio (PSPNR)~\cite{PSPNR}, a standard perceived quality metric.
It improves the classic Peak Signal-to-Noise Ratio (PSNR)~\cite{brain} by filtering out quality distortions that are imperceptible by users.
The key to PSPNR is the notion of {\em Just-Noticeable Difference} ({\em JND})~\cite{brain}, which is defined by the minimal changes in pixel values that can be noticed by viewers. 
PSPNR can be expressed as follows (Table~\ref{tab:terminology} summarizes the terminology):
\vspace{-0.1cm}
\begin{alignat}{2}\
\label{eq:pspnr} \PSPNR(\Quality) = 20 \times \log_{10}\frac{255}{\sqrt{\PMSE(\Quality)}}
\end{alignat}
\vspace{-0.3cm}
\begin{alignat}{2}\
\PMSE(\Quality)=\frac{1}{\Area}\sum_{i,j}\left[ |\Pixel_{i,j} - \hat{\Pixel}_{i,j}| - \JND_{i,j}\right]^2 \times \Delta (i, j)
\end{alignat}
\vspace{-0.3cm}
\begin{alignat}{2}\
\label{eq:delta} \Delta (i, j) =\left\{
\begin{aligned}
1, & &|\Pixel_{i,j} - \hat{\Pixel}_{i,j}| \ge \JND_{i,j} \\
0, & &|\Pixel_{i,j} - \hat{\Pixel}_{i,j}| < \JND_{i,j}
\end{aligned}
\right.
\end{alignat}
where $\Area$ denotes the image size, $\Pixel_{i,j}$ and $\hat{\Pixel}_{i,j}$ denote the pixel at $(i, j)$ of the original image and that of the image encoded at quality level $\Quality$ respectively, and $\JND_{i,j}$ denotes the JND at pixel $(i, j)$.

Intuitively, a change on a pixel value can affect the user-perceived quality (PSPNR) only if it is greater than the JND.
In other words, \camera{the notion of JND effectively provides an abstraction of users' sensitivities to quality distortion, which can be neatly incorporated in the quality metric of PSPNR.\questions{B.6}}

More importantly, we can incorporate the new quality-determining factors (\S\ref{subsec:opportunities}) by changing the calculation of JND---higher relative viewpoint moving speeds, greater DoF differences, or greater luminance changes will lead to higher JND.


%
%

\begin{table}[t]
\small
\begin{tabular}{r|l}
\hline
\textbf{Term} & \textbf{Brief description} \\ \hline\hline
$\Quality,\Chunk,\Tile$ & Quality level, chunk index, and tile index\\ \hline
$\Bitrate_{\Chunk,\Tile}(\Quality)$ & The bitrate of the $\Tile^{\textrm{th}}$ tile the $\Chunk^{\textrm{th}}$ chunk at quality $\Quality$\\ \hline
$\Pixel_{i,j},\hat{\Pixel}_{i,j}$ & Pixel value at $(i,j)$ on the original or encoded image \\ \hline
$\PSPNR(\Quality)$, $\PMSE(\Quality)$ & PSPNR (or PMSE~\cite{PSPNR}) of image at quality level $\Quality$ \\ \hline
$\JND_{i,j}$ & JND at pixel $i,j$ \\ \hline
$\Base_{i,j}$ & \begin{tabular}[c]{@{}l@{}}Content-dependent JND at pixel $(i,j)$: JND of zero \\ speed, luminance change, and DoF diff\end{tabular} \\ \hline
$\Action(x_1,x_2,x_3)$ & \begin{tabular}[c]{@{}l@{}}Action-dependent ratio: JND of speed $x_1$, luminance \\ change $x_2$, and DoF diff $x_3$, divided by $\Base$\end{tabular} \\ \hline
\end{tabular}
\vspace{0.3cm}
\tightcaption{Summary of terminology\vspace{-0.4cm}}
\label{tab:terminology}
\end{table}

\tightsubsection{Profiling JND of \vrvideos}
\label{subsec:jnd:details}

JND has been studied in the context of non-\vrvideos. 
However, prior work has focused on the impact of video content on JND. 
For instance, users tend to be less sensitive to quality distortion (\ie high JND) in areas of high texture complexity or excessively high/low luminance~\cite{PSPNR,distance,luminance1}.

As we have seen, however, \vrvideos are different, in that a user's sensitivity may vary with the viewpoint movement as well. 
In other words, 
the JND of a pixel $(i,j)$ is also dependent on the following values: (1) the speed $v$ of an object $O$ (of which pixel $(i,j)$ is a part) relative to the viewpoint;
(2) the luminance $l$ of $O$ relative to where the viewpoint focused on 5 seconds ago; 
(3) the DoF difference $d$ between $O$ and the viewpoint focused object;
and (4) the base JND $\Base_{i,j}$, defined by the JND when there is no viewpoint movement (\ie $v=0,l=0$) or DoF difference ($d=0$).
Because $\Base_{i,j}$ is only dependent on the video content, we refer to it as the {\em content-dependent JND}. We calculate $\Base_{i,j}$ using the same JND formulation from the prior work~\cite{PSPNR,distance}.

To quantify the impact of $v,l,d$ on JND, we ran a user study using a similar methodology to the prior studies~\cite{distance,PSPNR}.
\camera{Readers can find more details of our methodology in Appendix.
The study has 20 participants. 
Each participant is asked to watch a set of 43 short test videos, each generated with a specific level of quality distortion. 
The quality distortion is gradually increased until the participant reports that the distortion becomes noticeable.
\questions{A.2}\questions{A.5}\questions{D.2}}

\begin{figure}
  \centering
    \begin{minipage}[t]{0.335\linewidth}
\centering
\includegraphics[width=\linewidth]{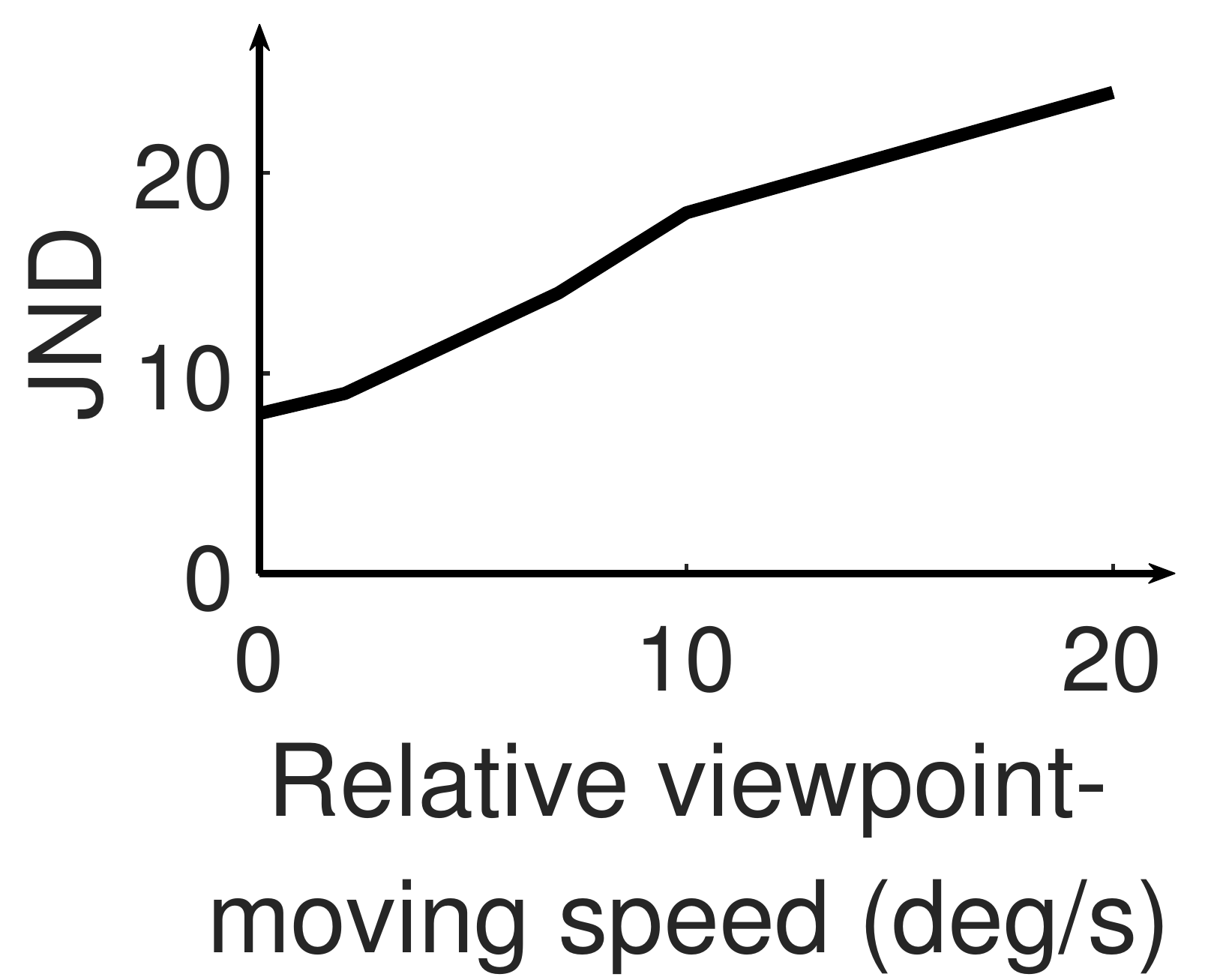}
\label{fig:side:a}
\end{minipage}\hspace{-0.15cm}
   \begin{minipage}[t]{0.335\linewidth}
\centering
\includegraphics[width=\linewidth]{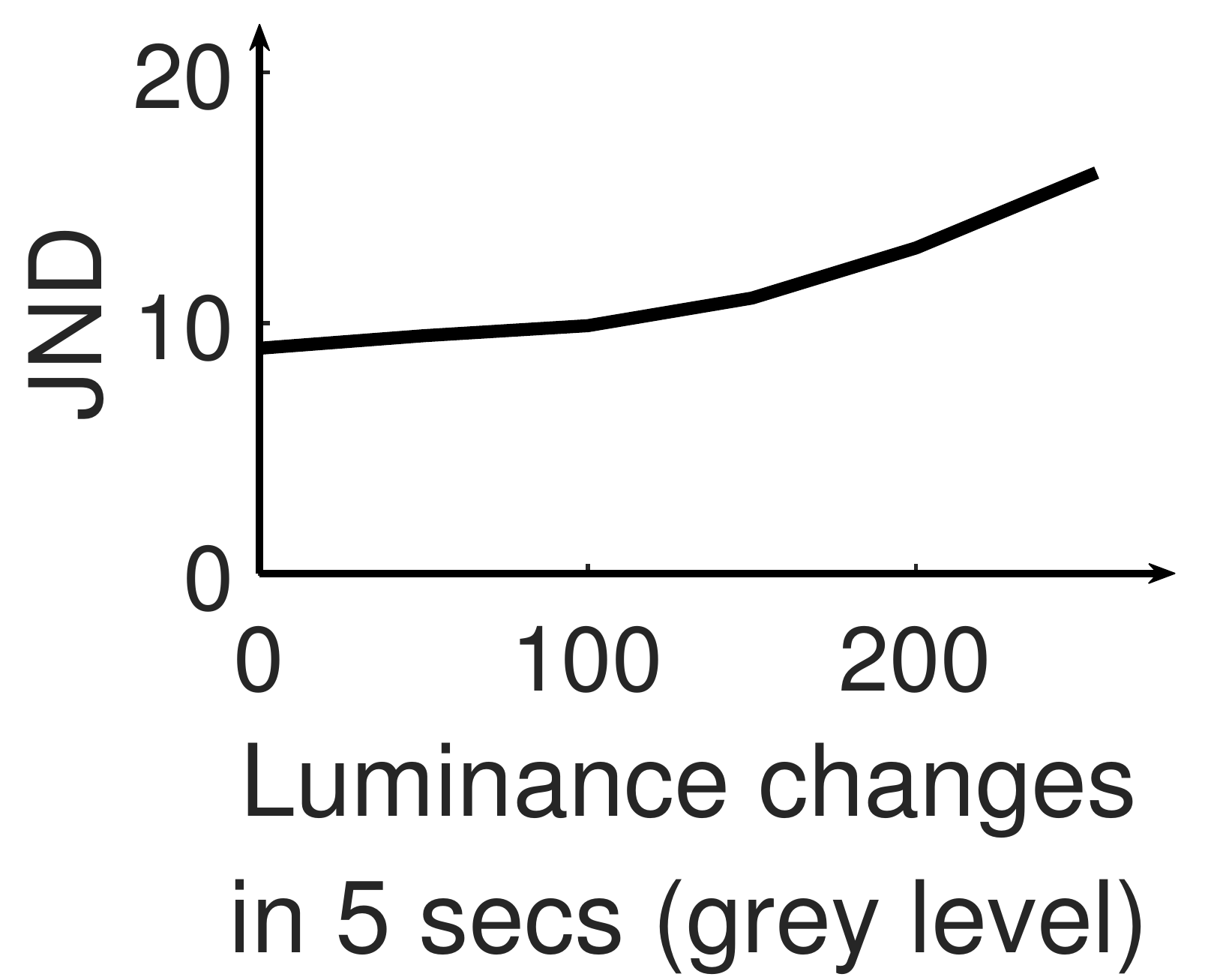}
\label{fig:side:a}
\end{minipage}\hspace{-0.1cm}
  \begin{minipage}[t]{0.335\linewidth}
\centering
\includegraphics[width=\linewidth]{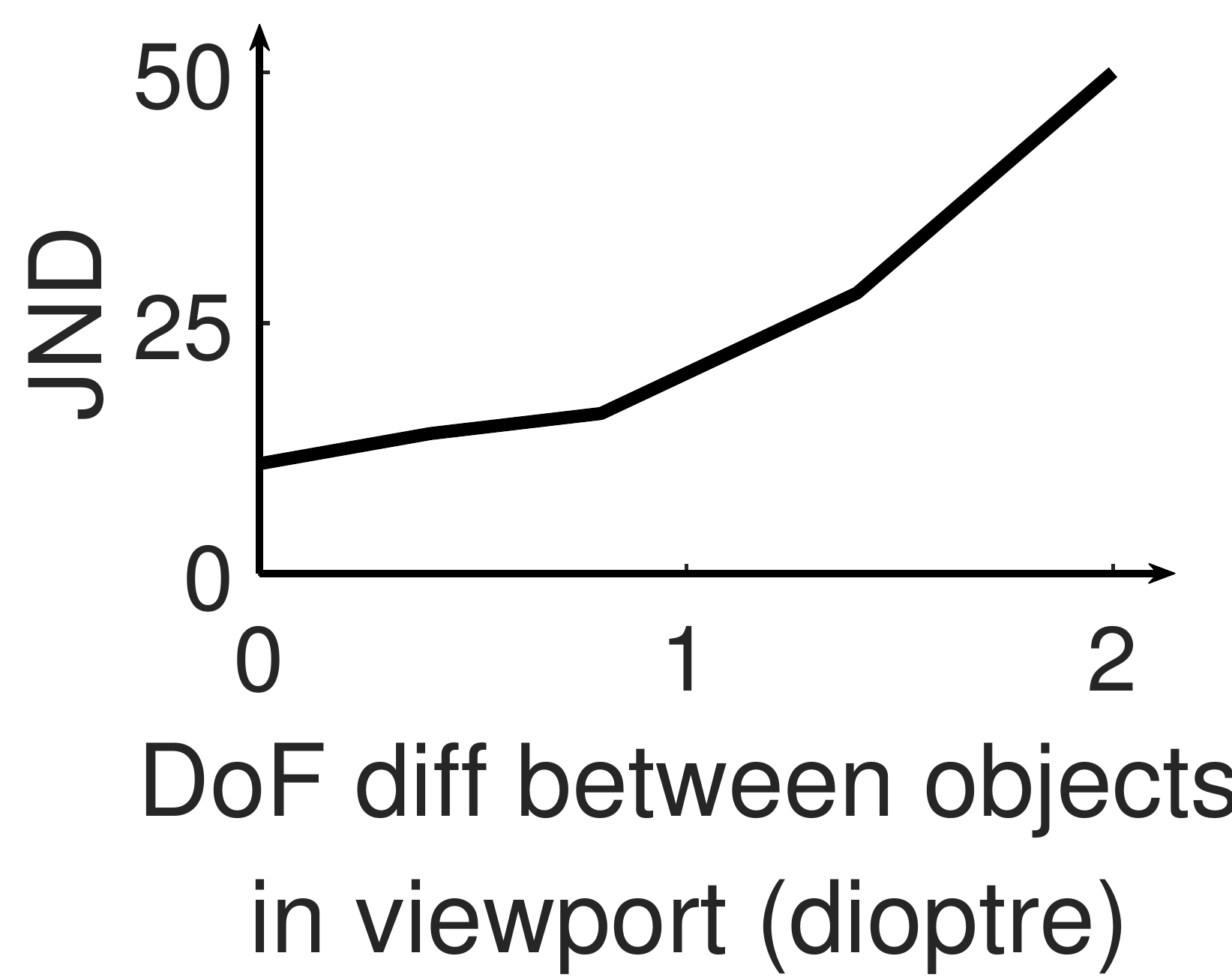}
\label{fig:side:a}
\end{minipage}
  \vspace{-0.3cm}
  \tightcaption{Impact of individual factors on JND. 
  \vspace{-0.3cm}
  }
  \label{fig:single-factor}
 \end{figure}

\mypara{Impact of individual factors}
Figure~\ref{fig:single-factor} shows how JND changes with the relative viewpoint-moving speed, luminance change, or DoF difference, while the other two factors are kept to zero.
As expected, JND increases (\ie users become less sensitive to quality distortion) monotonically with higher relative viewpoint-moving speeds, greater luminance changes, or sharper DoF differences.
Formally, we use $F_v(x)$ ($F_l(x)$ or $F_d(x)$) to denote the ratio between the JND when $v=x$ ($l=x$ or $d=x$) and the JND when $v=0$ ($l=0$ or $d=0$), while holding the other two factors $l,d$ at zero. 
We call $F_v(x)$, $F_l(x)$, and $F_d(x)$ the {\em viewpoint-speed multiplier}, the {\em luminance-change multiplier}, and the {\em DoF-difference multiplier}, respectively.

\begin{figure}
  \centering
     \begin{minipage}[t]{0.5\linewidth}
\centering
\includegraphics[width=\linewidth]{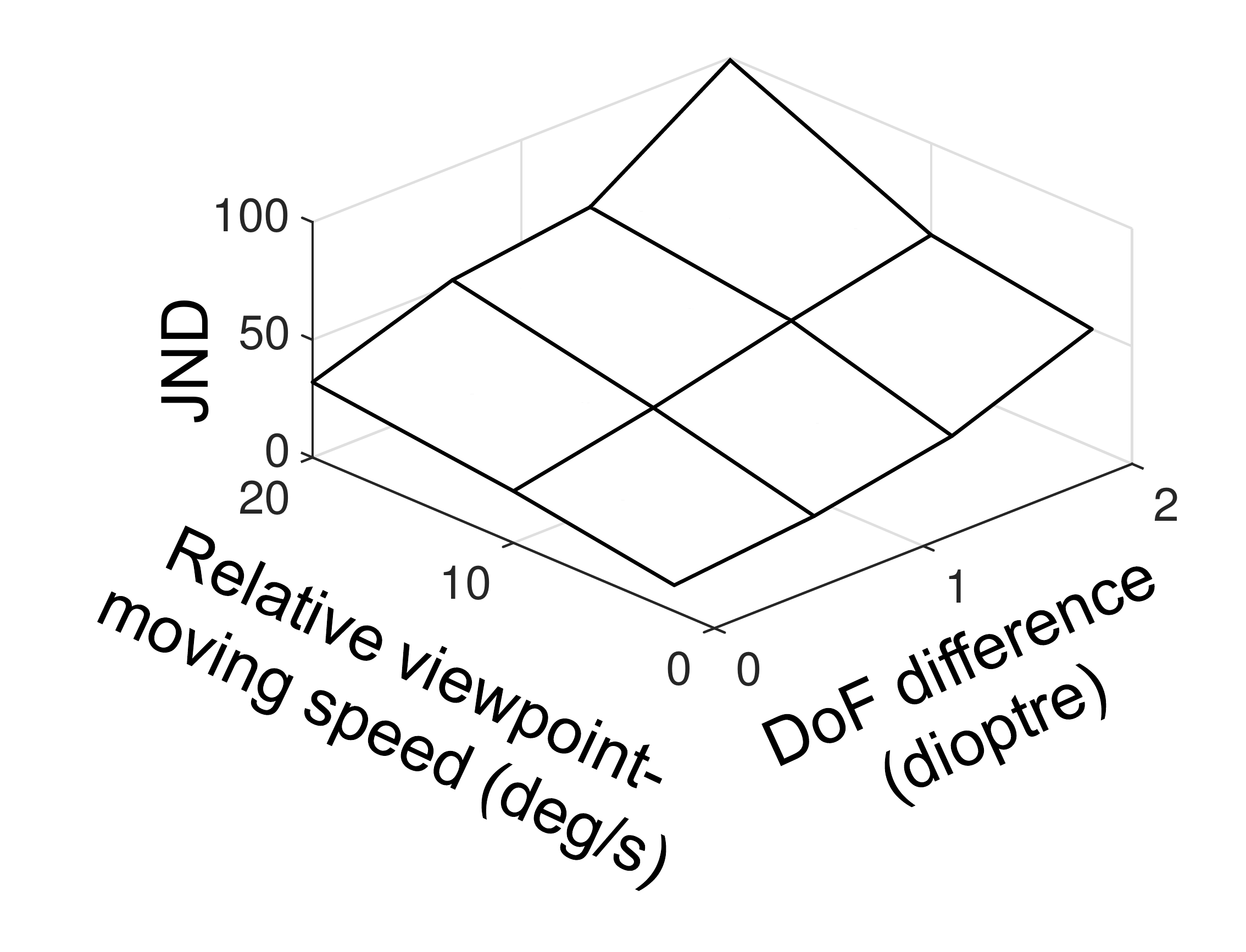}
 {\small (a) JND vs. viewpoint-moving\\ speed \& DoF difference}
\label{fig:side:a}
\end{minipage}\hspace{-0.25cm}
   \begin{minipage}[t]{0.5\linewidth}
\centering
\includegraphics[width=\linewidth]{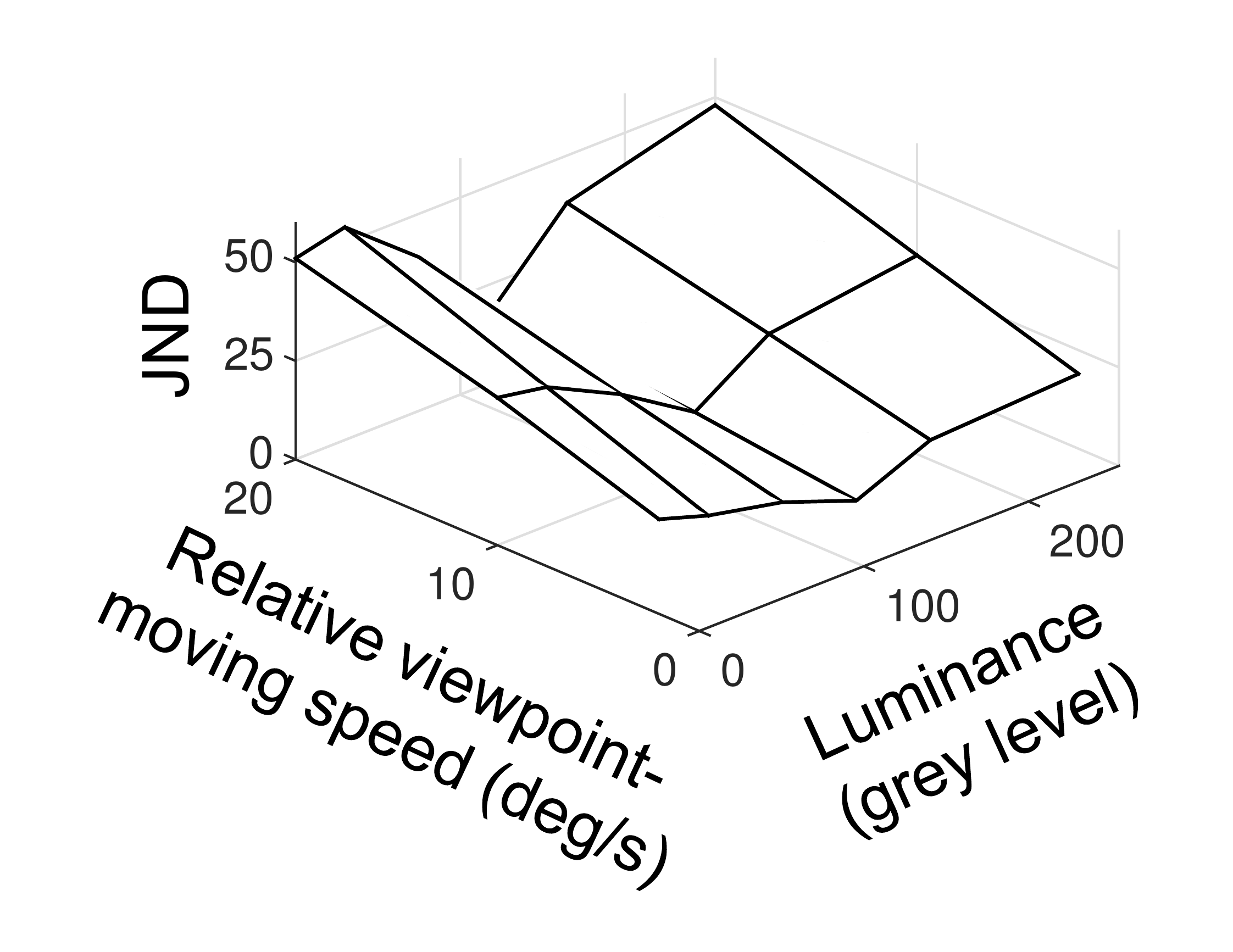}
 {\small (b) JND vs. viewpoint-moving \\speed \& luminance}
\label{fig:side:a}
\end{minipage}\hspace{0.5cm}
  \tightcaption{Joint impact of two factors on JND.\vspace{-0.2cm}}
  \label{fig:two-factor}
 \end{figure}

\mypara{Impact of multiple factors} 
Figure~\ref{fig:two-factor} shows the joint impact of two factors on JND. 
In Figure~\ref{fig:two-factor}(a), we notice that JND under viewpoint-moving speed $v=x_1$ and DoF difference $d=x_2$ can be approximated by the product of $\Base\cdot F_v(x_1)\cdot F_d(x_2)$, where $\Base$ is the content-dependent JND (\ie when $v=0,l=0,d=0$).
This suggests the impact of these two factors on JND in this test appears to be {\em independent}.
We see similar statistical independence between the impact of luminance change and that of viewpoint-moving speed or DoF difference.
Figure~\ref{fig:two-factor}(b) shows the joint impact of viewpoint-moving speed (one of the \vrvideo-specific factors) and the viewpoint's current luminance value (one of the traditional factors that affect JND). 
The figure also shows the impact of these two factors on JND in this test appears to be independent.
Notice that the impact of current luminance value on JND is non-monotonic because quality distortion tends to be less perceptible when the video is too bright or too dark.

The observation that different \vrvideo-specific factors appear to have independent impact on JND is well aligned with previous findings that other factors (\eg content luminance, distance-to-viewpoint) have largely independent impact on JND~\cite{distance,PSPNR,brain}.


\mypara{Putting it together}
Now, we define a new way of calculating JND for \vrvideos, called {\em \vrjnd}, as follows:\questions{A.6}
\begin{alignat}{2}
\vspace{-0.1cm}
\JND_{i,j} = \Base_{i,j} \cdot F_{v}(x_1) \cdot F_{d}(x_2) \cdot F_{l}(x_3) \triangleq \Base_{i,j} \cdot \Action(x_1,x_2,x_3)
\vspace{-0.1cm}
\label{equ:jnd}
\end{alignat}
In other words, \vrjnd is the product of the {\em content-dependent JND}, and the {\em action-dependent ratio} $\Action(x_1,x_2,x_3)$,  which is the product of the viewpoint-speed multiplier, luminance-change multiplier, and DoF-difference multiplier.
As we will see in \S\ref{subsec:compatible}, this separation has a great implication that the content-dependent JND can be pre-calculated, whereas the action-dependent ratio can be determined only in realtime, without the help from the server.

\begin{figure}
  \centering
  \includegraphics[width=0.3\textwidth]{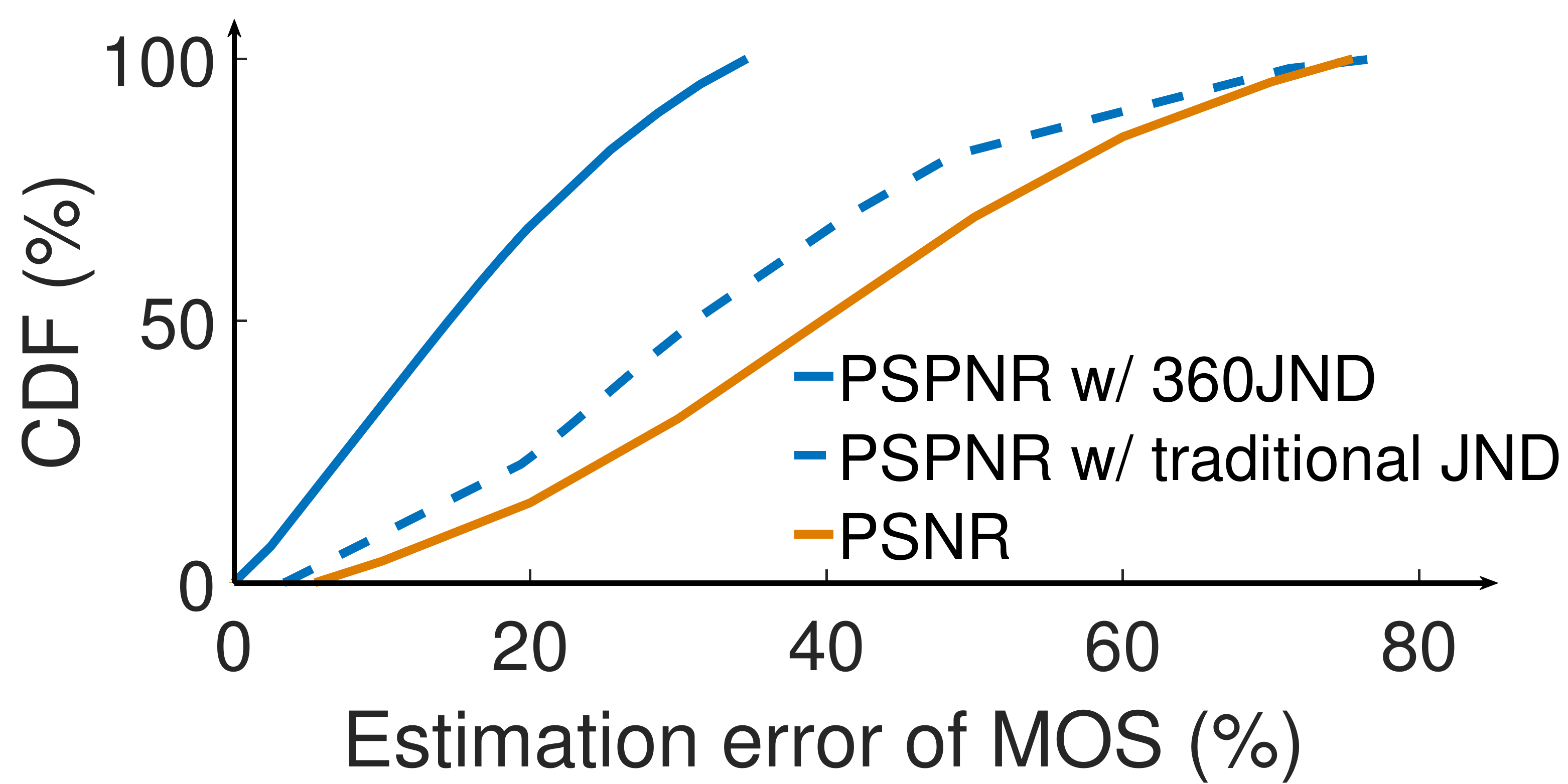}
  \tightcaption{\vrjnd-based PSPNR can estimate MOS much more accurately than the traditional PSPNR and PSNR.}
  \label{fig:pspnr-accuracy}
 \end{figure}

\mypara{Validation of usefulness}
To verify the usefulness of the new \vrjnd model, we plug the \vrjnd in the PSPNR calculation in Equation~\ref{eq:pspnr}-\ref{eq:delta}, and then check how well the resulting PSPNR value correlates with the actual user rating (MOS) from 20 participants over 21 \vr videos.
(See \S\ref{subsec:eval:methodology} for more details on how user rating is recorded.)
For each video, we calculate the average \vrjnd-based PSPNR across users as well as the MOS.
Then, we build a linear predictor that estimates MOS based on average PSPNR.
As reference points, we similarly build a linear predictor using traditional JND-based PSPNR and a predictor using PSNR (JND-agnostic).
Figure~\ref{fig:pspnr-accuracy} shows the distribution of relative estimation errors of the three predictors ($\frac{|MOS_{predict}-MOS_{real}|}{MOS_{real}}$).
We see that \vrjnd-based PSPNR can predict MOS much more accurately than the alternatives, which suggests the three \vrvideo-specific factors have a strong influence on \vrvideo perceived quality.

%% file: tiling.tex

\tightsection{\name: Video tiling}
\label{sec:tiling}

\begin{figure}
  \centering
  \includegraphics[width=0.45\textwidth]{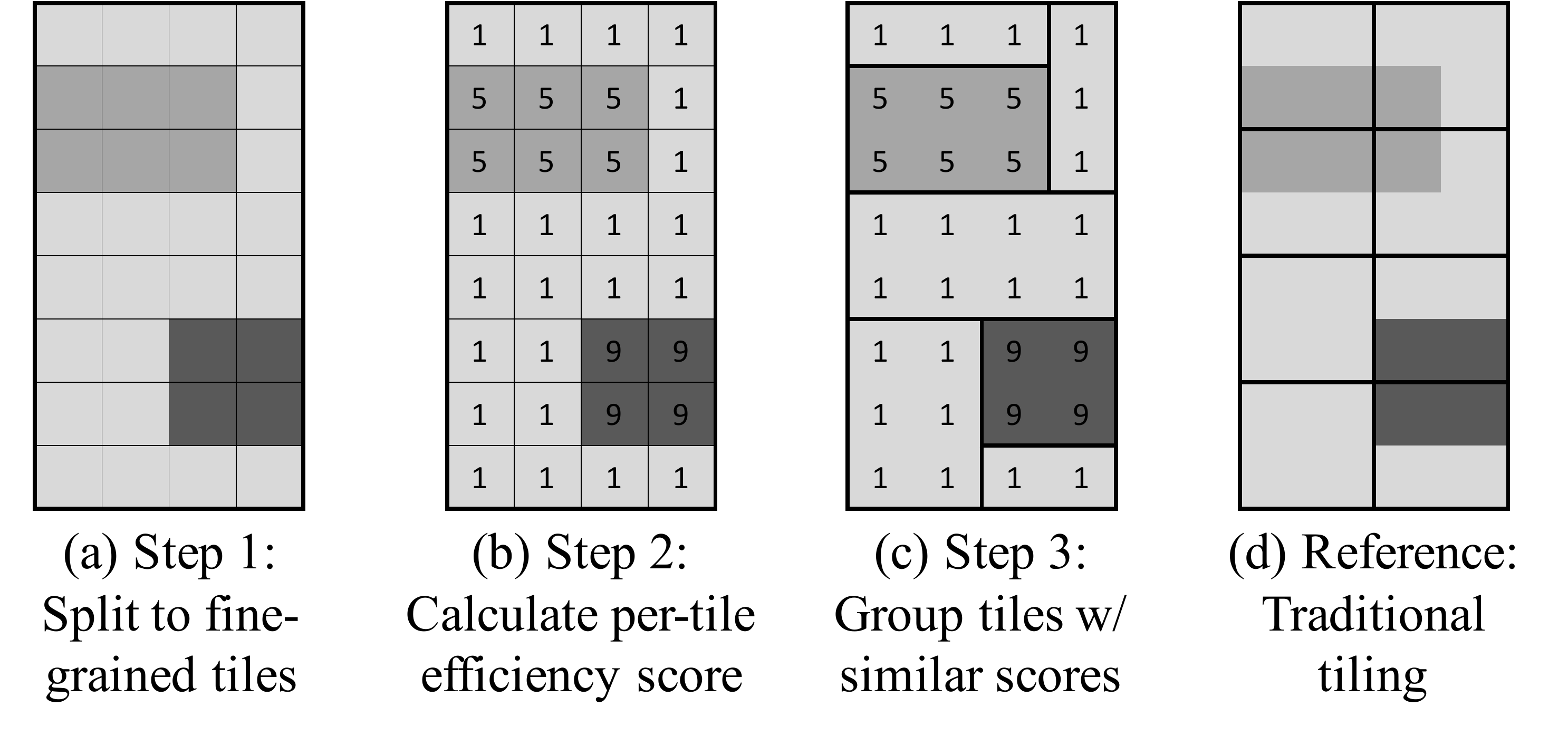}
  \tightcaption{The steps of \name tiling. The shades indicate regions with similar efficiency score.
  \vspace{-0.1cm}
  }
  \label{fig:tiling-steps}
\end{figure}

Next, we describe \name's tiling scheme, which leverages the quality metric introduced in \S\ref{sec:jnd}.
Like other DASH-based videos, \name first chops a \vrvideo into chunks of equal length (\eg one second), and then spatially splits each chunk into tiles by the following steps (as illustrated in Figure~\ref{fig:tiling-steps}). 
\camera{}

\myparashort{Step~1: Chunking and fine-grained tiling.}
\name begins by splitting each chunk into fine-grained square-shape {\em unit tiles} with a 12-by-24 grid.
Each unit tile is a video clip containing all content within the square-shape area in the chunk's duration. 
These unit tiles are the building blocks which \name then groups into coarser-grained tiles as follows.


\myparashort{Step~2: Calculating per-tile efficiency scores.}
Then \name calculates an {\em efficiency score} for each unit tile, which is defined by how fast the tile's quality grows with the quality level. 
Formally, the efficiency score of unit tile $\Tile$ of chunk $\Chunk$ is 
\begin{alignat}{2}
\vspace{-0.2cm}
\Efficiency_{\Chunk,\Tile}=\frac{\PSPNR_{\Chunk,\Tile}(\Quality_{high})-\PSPNR_{\Chunk,\Tile}(\Quality_{low})}{\Quality_{high}-\Quality_{low}}
\vspace{-0.2cm}
\label{equ:efficiency}
\end{alignat}
where $\PSPNR_{\Chunk,\Tile}(\Quality)$ is the PSPNR (perceived quality calculated by Equation~\ref{eq:pspnr}) of the unit tile when it is encoded at quality level $\Quality$; and $\Quality_{low}$ (and $\Quality_{high}$) denotes the lowest (and highest) quality level. 
There are two caveats. 
\camera{First, we assume the PSPNR of a unit tile is known. 
We will explain how to estimate them offline at the end of this section.\questions{A.7}}
Second, Equation~\ref{equ:efficiency} assumes that $\PSPNR$ grows linearly with $\Quality$.
This may not be true, but we found this assumption is a good approximation, and our solution does not crucially rely on it. 
We leave further refinements for future work.

\myparashort{Step~3: Tile grouping.}
Finally, \name groups the 12$\times$24 unit tiles into $N$ (by default, 30) variable-size coarse-grained rectangle tiles, which will eventually be used by \name to encode the video. 
The goal of this grouping process is to reduce the variance of efficiency scores among the unit tiles in the same group (coarse-grained tile). 
More specifically, we try to minimize the weighted sum of these variances, where each variance is weighted by the area of the group. 
The intuition is that, 
because a higher/lower efficiency score means a tile will produce higher/lower PSPNR at the same quality level, the tiles with similar efficiency scores tend to be assigned with similar quality levels during playback, so grouping these unit tiles will have limited impact on quality adaptation.
At the same time, having fewer tiles can significantly reduce the video size, as it avoids re-encoding the boundaries between small tiles.

Our grouping algorithm starts with one hypothetical rectangle that includes all 12$\times$24 unit tiles (\ie the whole \vrvideo).
It then uses a top-down process to enumerate many possible ways of partitioning this hypothetical rectangle into $N$ rectangles, each representing a coarse-grained tile.
It begins by splitting this hypothetical rectangle into two rectangles along each possible vertical or horizontal boundary.
Then it iteratively picks one of the existing rectangles that has more than one unit tile, and then similarly splits it, vertically or horizontally, into two rectangle tiles.
This process runs repeatedly until there are $N$ rectangles (coarse-grained tiles). 
This process is similar to how the classic 2-D clustering algorithm~\cite{2dclustering} enumerates the possible partitions of a 2D space. 

\mypara{Calculating efficiency scores offline}
We assume each video has some history viewpoint trajectories, like in~\cite{ClusTile,CLS}. 
For each tile, we compute the PSPNR under each history viewpoint trajectory, average the PSPNRs per tile across all trajectories, and derive the efficiency score per tile using Equation~\ref{equ:efficiency}.
\camera{The resulting PSPNR per tile takes both content information {\em and} viewpoint movements into account.
Once the tiles are determined offline, \name does not adjust them during playback, so the video does not need to be re-encoded.
\questions{B.2}}
\camera{We acknowledge that computing PSPNR with the average history viewpoint movements might cause suboptimal quality for users with atypical viewing behaviors.
That said, we found that the lowest perceived quality across the users in our traces is at most 10\% worse than the mean quality (Figure~\ref{fig:robust-1}(b)).
\questions{A.4}\questions{A.7}}

%% file: control.tex
\tightsection{\name: Quality adaptation}
\label{sec:control}

The design of \name's quality adaptation logic addresses two following questions. 
(1) How to adapt quality in the presence of noisy viewpoint estimates (\S\ref{subsec:adaptation})?
And (2) how to be deployable on the existing DASH protocol (\S\ref{subsec:compatible})?

\tightsubsection{Robust quality adaptation}
\label{subsec:adaptation}
\name adapts quality at both the {\em chunk} level and the {\em tile} level.
First, \name uses MPC~\cite{MPC} to determine the bitrate of each chunk, to meet buffer length target under the predicted bandwidth.
The chunk's bitrate determines the total size of all tiles in the chunk.

Then, within the chunk $\Chunk$, \name determines the quality level $\Quality_\Tile$ of each tile $\Tile\in\{1,\dots,N\}$ ($N$ is the number of tiles per chunk), to maximize the overall perceived quality (PSPNR) while maintaining total size of the tiles below the chunk's bitrate $\bitrate_\Chunk$. 
According to Equation~\ref{eq:pspnr}, the overall PSPNR of the $N$ tiles is
$\PSPNR=20\times\log_{10}\frac{255}{\sqrt{\PMSE}}$, where $\PMSE=(\sum_{\Tile=1,\dots,N}\Area_\Tile\cdot\PMSE_\Tile(\Quality_\Tile))/(\sum_{\Tile=1,\dots,N}\Area_\Tile)$, and $\Area_\Tile$ is the area size of tile $\Tile$. 
Since the total area of all tiles is constant, the tile-level quality allocation can be formulated as follows:
\begin{align*}
\textrm{min} & \sum_{\Tile=1,\dots,N}\Area_\Tile\cdot\PMSE_\Tile(\Quality_\Tile)&\textrm{/* Maximizing overall PSPNR */}\\
\textrm{s.t.} & \sum_{\Tile=1,\dots,N}\Bitrate_{\Chunk,\Tile}(\Quality_\Tile)\leq \bitrate_\Chunk &\textrm{/* Total tile size $\leq$ chunk bitrate */}
\end{align*}
To solve this optimization problem, we enumerate the possible assignment of 5 quality levels in each of the $N$ tiles, but instead of an exhaustive search (which has $5^N$ outcomes), \name prunes the search space using the following observation.
For any pair of tiles ($\Tile_1$ and $\Tile_2$), if we found one quality assignment (\eg assigning $\Quality_1$ to $\Tile_1$ and $\Quality_2$ to $\Tile_2$) is ``strictly'' better (\ie producing higher PSPNR {\em and} smaller total tile size) than another assignment (\eg assigning $\Quality_3$ to $\Tile_1$ and $\Quality_4$ to $\Tile_2$), then we can safely exclude the latter assignment when iterating the quality assignments of the remaining tiles.

\begin{figure}
 \centering
 \includegraphics[width=2.5in]{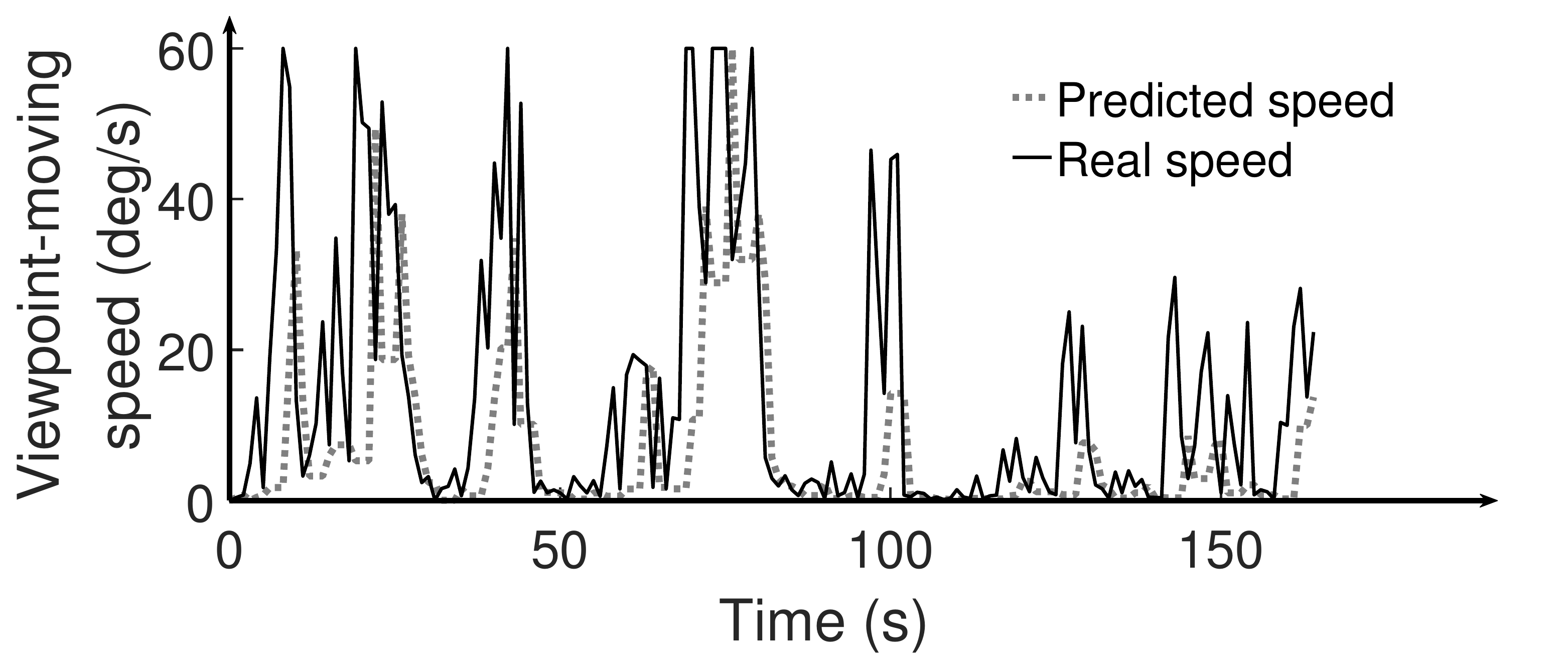}
 \tightcaption{\name can reliably estimate a lower bound (dotted line) of the actual viewpoint-moving speed (solid line), which is often sufficient for accurate PSPNR estimation.
 \vspace{-0.2cm}}
 \label{speed_analysis}
 \end{figure}

\camera{
\mypara{Coping with viewpoint estimation errors} 
In theory, optimal quality adaptation requires accurate PSPNR estimation, which relies heavily on accurate estimation of viewpoint-moving speeds, DoF differences, and luminance changes.
In practice, however, we found that predicting an approximate {\em range} of these three factors is sufficient to inform optimal quality selection. 
The reason is two-fold. 
On one hand, if the head has little or slow movement (\eg staring at an object), it is trivial to accurately predict the viewpoint-moving speed, DoF, and luminance.
On the other hand, if the viewpoint moves arbitrarily, it is difficult to predict the exact viewpoint-moving speed, DoF, and luminance, but it is still plausible to estimate a {\em lower bound} for each factor using recent history.
For instance, the lowest speed in the last two seconds serves a reliable conservative estimator of the speed in the next few seconds (Figure~\ref{speed_analysis}).
Although these lower bounds would lead \name to make conservative decisions (\eg assigning a higher-than-necessary quality), these conservative decisions still bring sizable improvement over the baselines which completely ignore the impact of viewpoint-moving speed, DoF, and luminance. \questions{D.1}}

\tightsubsection{DASH-compatible design}
\label{subsec:compatible}

While the logical workflow of \name is straightforward, it is incompatible with the popular DASH protocol~\cite{DASH}. 
This is because \name's quality adaptation is based on PSPNR (Equation~\ref{eq:pspnr}), 
which requires both viewpoint movements (only available on the client) and the pixels of the video content (only available on the server). 
\camera{This, however, violates the key tenet of the popular DASH protocol that servers must be passive while clients adapt bitrate locally without aid of the server.\questions{B.8}}

\begin{figure}
  \centering
  \includegraphics[width=0.5\textwidth]{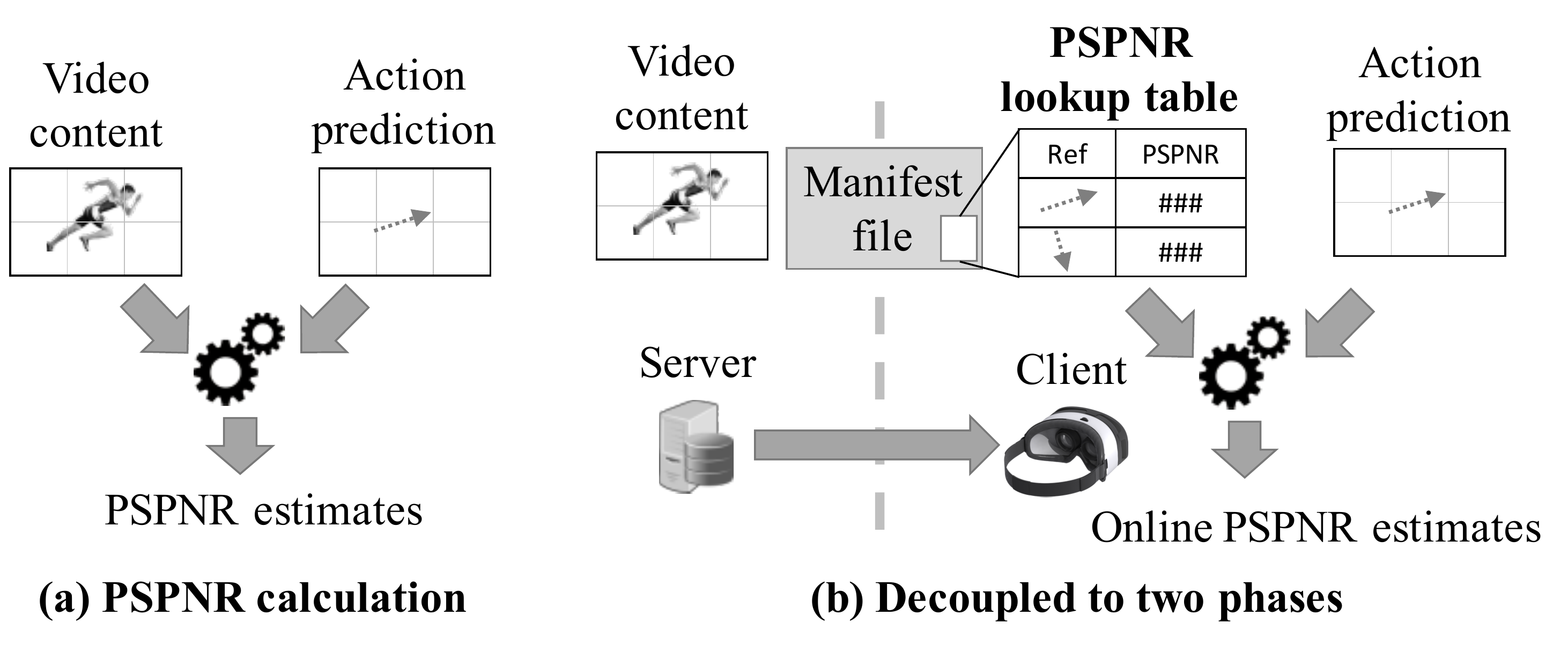}
  \vspace{-0.4cm}
  \tightcaption{
  (a) \name calculates PSPNR by first pre-processing content-dependent information offline, and then combining it with online viewpoint predictions by the client.
  (b) The offline content-dependent information (represented by PSPNR lookup table) is included in the manifest file sent to the client at the beginning of a video.\vspace{-0.2cm}
  }
  \label{fig:decoupling}
  \end{figure}

Fortunately, \name can be implemented in a way that is compatible with the DASH protocol. 
The basic idea is to decouple the calculation of PSPNR into two phases (as illustrated in Figure~\ref{fig:decoupling}).
In the offline phase, the video provider pre-calculates the PSPNR for some ``representative'' viewpoint movements and stores them in a {\em PSPNR lookup table}.
In particular, we choose $n$ representative values for each of the viewpoint speed, DoF difference and luminance change, which produces $n^3$ combinations and the corresponding PSPNR values in the lookup table. 
The PSPNR lookup table is sent to the client as part of the DASH manifest file at the beginning of a video.
In the online phase, the client uses the PSPNR lookup table to estimate PSPNR under the actual viewpoint movement.


\tightsubsection{System optimization}
\label{subsec:optimization}

\begin{figure}[t]
  \centering
  \includegraphics[width=0.48\textwidth]{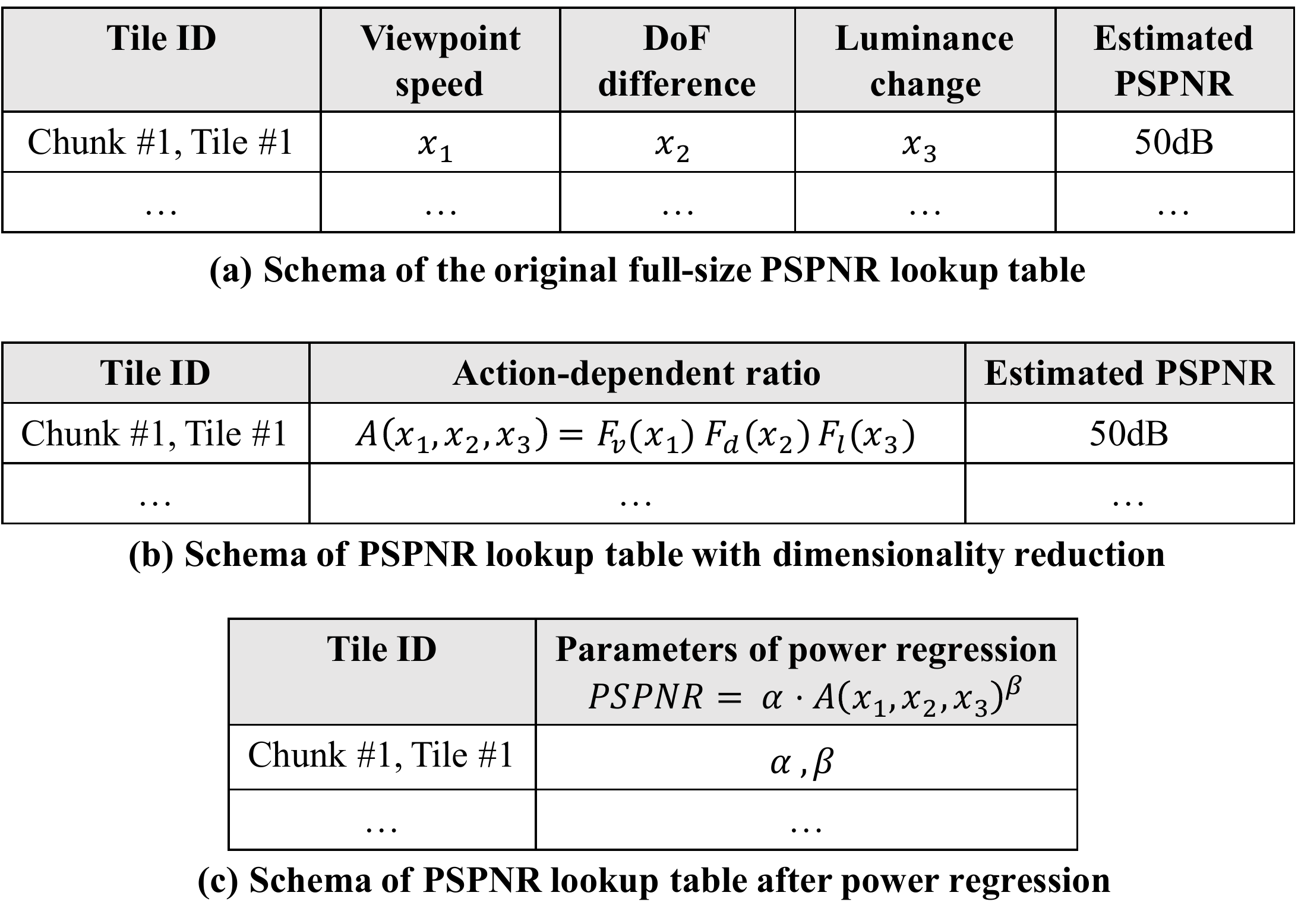}
  \vspace{-0.4cm}
  \tightcaption{The schema of PSPNR lookup table.\vspace{-0.1cm}}
\label{fig:pspnr-table-compression}
\end{figure}

\camera{\mypara{Compressing PSPNR lookup table}
A PSPNR lookup table (see Figure~\ref{fig:pspnr-table-compression}(a) for an example) includes, for each tile, the PSPNR estimates of every possible combination of viewpoint-moving speed, luminance change, and DoF difference. 
Without compression, the PSPNR lookup table can be 10~MB for a 5-minute video, which can significantly inflate the manifest file size. 
We address this problem by two techniques, which produce an approximate yet more compressed representation of the PSPNR lookup table.
First, we reduce the PSPNR lookup table from ``multi-dimensional'' to ``one-dimensional'' (Figure~\ref{fig:pspnr-table-compression}(b)), by replacing the viewpoint speed, DoF, luminance with the products of their multipliers (defined in \S\ref{subsec:jnd:details}).
Using their products, \ie the action-dependent ratios (see Equation~\ref{equ:jnd}) to index the PSPNR lookup table, we can avoid enumerating a large number of combinations of viewpoint speed, DoF, and luminance.
Second, instead of keeping a map between action-dependent ratios and their corresponding PSPNR of each tile, we found that their relationship in a given tile can be interpolated by a power function. 
Thus, we only need two parameters to encode the relationship between PSPNR and action-dependent ratio (Figure~\ref{fig:pspnr-table-compression}(c)).
With these optimizations, we can compress the manifest file from 10~MB to $\sim$50~KB for a 5-minute video.
\questions{A.8}}

\mypara{Reducing PSPNR computation overhead}
Per-frame PSPNR calculation, in its original form (Equation~\ref{eq:pspnr}), can be $\sim50\%$ slower than encoding the same video. 
To reduce this overhead, we extract one frame from every ten frames and use its PSPNR as the PSPNR of other nine frames.
This saves the PSPNR computation overhead by 90\%, and we found this is as effective as per-frame PSPNR computation. 

%% file: impl.tex

\tightsection{Implementation}
\label{sec:impl}

Here, we describe the changes needed to deploy \name in a DASH video delivery system.
We implement a prototype of \name with 15K lines of codes by C++, C\#, Python, and Matlab~\cite{pano}.


 
\mypara{Video provider}
The video provider preprocesses a \vrvideo in three steps.
First, we extract features from the video, such as object trajectories, content luminance, and DoF, which are needed to calculate the PSPNR of the video under each of history viewpoint movements.
In particular, to detect the object trajectories, we use Yolo\cite{yolov3} (a neural network-based multi-class object detector) to detect objects in the first frame of each second, and then use a tracking logic~\cite{Kcf} to identify the trajectory of each detected object in the remaining of the second. 
Then we temporally split the video into 1-second chunks, use the tiling algorithm described in \S\ref{sec:tiling} to spatially split each chunk into $N$ (by default, 30) tiles using FFmpeg CropFilter~\cite{ffmpeg}, and encode each tile in $5$ QP levels (\eg \{22, 27, 32, 37, 42\}). 
Finally, we augment the video's manifest file with additional information. 
In particular, each tile includes the following information (other than available quality levels and their corresponding URLs):
(1) the coordinate of the tile's top-left pixel (this is needed since the tiles in \name may not be aligned across chunks);
(2) average luminance within the tile;
(3) average DoF within the tile;
(4) the trajectory of each visual object (one sample per 10 frames); and
(5) the PSPNR lookup table (\S\ref{subsec:optimization}).

\mypara{Video server}
Like recent work on \vrvideos~\cite{Flare}, \name does not need to change the DASH video server.
It only needs to change the client-side player as described next. 




\mypara{Client-side adaptation}
We built \name client on a FFmpeg-based~\cite{ffmpeg} mockup implementation of the popular dash.js player~\cite{DASH}.
To let the client use \name's quality adaptation logic, we make the following changes.
First, the player downloads and parses the manifest file from the server. 
We change the player's manifest file parser to extract information necessary for \name's quality adaptation. 
Second, we add the three new functionalities to the DASH bitrate adaption logic.
The {\em viewpoint estimator} predicts viewpoint location in the next 1-3 seconds, using a simple linear regression over the recent history viewpoint locations~\cite{Flare,Qian}.
Then the {\em client-side PSPNR estimator} compares the predicted viewpoint movements with the information of the tile where the predicted viewpoint resides (extracted from the manifest file) to calculate the relative viewpoint speed, the luminance change, and the DoF difference. 
These factors are then converted to the PSPNR of each tile in the next chunk using the PSPNR lookup table (\S\ref{subsec:compatible}).
Finally, after the DASH bitrate adaptation algorithm~\cite{MPC} decides the bitrate of a chunk, the {\em tile-level bitrate allocation logic} assigns quality levels to its tiles using the logic described in \S\ref{subsec:adaptation}.

\mypara{Client-side streaming}
We fetch the tiles of each chunk as separate HTTP objects (over a persistent HTTP connection), then decode these tiles in parallel into separate in-memory YUV-format objects using FFmpeg, and finally stitch them together into a panoramic frame using in-memory copy.
We use the coordinates of each tile (saved in the manifest file) to decide its location in the panoramic frame. 
As an optimization, the copying of tiles into a panoramic frame can be made efficient if the per-tile YUV matrices are copied in a row-major manner (\ie which is aligned with how matrices are laid out in memory), using the compiler-optimized \texttt{memcpy}.
As a result, the latency of stitching one panoramic frame is 1ms.

%% file: eval.tex
\tightsection{Evaluation}
\label{sec:eval}

We evaluate \name using both a survey-based user study and trace-driven simulation. 
Our key findings are the following.

\begin{packeditemize}

\item Compared to the state-of-the-art solutions, \name improves perceived quality without using more bandwidth: 25\%-142\% higher mean opinion score (MOS) or 10\% higher PSPNR with the same or less buffering across a variety of \vrvideo genres.

\item \name achieves substantial improvement even in the presence of viewpoint/bandwidth prediction errors. 

\item \name imposes minimal additional systems overhead and reduces the resource consumption on the client and the server.


\end{packeditemize}


\tightsubsection{Methodology}
\label{subsec:eval:methodology}
\mypara{Dataset}
We use $50$ \vrvideos (7 genres and $200$ minutes in total).
Among them, $18$ videos (also used in \S\ref{subsec:potentials}) have actual user viewpoint trajectories from a set of 48 users (age between 20 and 26). 
Each viewpoint trajectory trace is recorded on an HTC Vive~\cite{vive} device. \camera{The viewpoints are refreshed every 0.05s, which is typical to other mainstream VR devices~\cite{iQiyi,Youku,vive}. \questions{C.8}}
Each video is encoded into $5$ quality levels (QP=$22,27,32,37,42$) and 1-second chunks using the x264 codec~\cite{X264}. Table~\ref{tab:dataset} gives a summary of our dataset.

\mypara{Baselines}
We compare \name with two recent proposals, Flare~\cite{Flare} and ClusTile~\cite{ClusTile}.
They are both viewport-driven, but they prioritize the quality within the viewport in different ways. 
Flare uses the viewport location to spatially allocate different quality to the uniform-size tiles, whereas ClusTile uses the viewport to determine the tile shapes. 
Conceptually, \name combines their strengths by extending both tiling and quality allocation using the new \vrvideo quality model.
As a reference point, we also consider the baseline that streams the whole video in its 360\textdegree\xspace view.
For a fair comparison, all baselines and \name use the same logic for viewpoint prediction (linear regression) and chunk-level bitrate adaptation~\cite{MPC}.


\begin{table}[t]
\begin{small}
\begin{tabular}{rl}
\hline
{\bf Total \# videos}       & 50 (18 with viewpoint traces of 48 users)              \\ \hline
{\bf Total length (s)}    & 12000                                          \\ \hline
{\bf Full resolution}       & 2880 x 1440                                     \\ \hline
{\bf Frame rate}            & 30                                          \\ \hline
{\bf Genres (\%)} &
\begin{tabular}[c]{@{}l@{}}Sports (22\%), Performance (20\%), \\ Documentary (14\%), other(44\%)
\end{tabular} \\ \hline
\end{tabular}
\end{small}
\vspace{0.3cm}
\tightcaption{Dataset summary 
}
\vspace{-0.2cm}
\label{tab:dataset}
\end{table}

\begin{table}[t]
\begin{small}
\begin{tabular}{cccccc}
\hline
{\bf PSPNR (\vrjnd-based)}       &$\leq45$ &46-53 &54-61 &62-69 &$\geq70$            \\ \hline
{\bf MOS}      &1 &2   & 3  & 4   & 5                  \\ \hline
\end{tabular}
\end{small}
\vspace{0.3cm}
\tightcaption{Map between MOS and new \vrjnd-based PSPNR (\S\ref{sec:jnd})
}
\vspace{-0.6cm}
\label{tab:pspnr-mos}
\end{table}

\mypara{Survey-based evaluation}
We run a survey-based evaluation on 20 participants. 
Each participant is asked to watch 7 videos of different genres, each played in 4 versions: 2 methods (\name and Flare) and 2 bandwidth conditions (explained in next paragraph). 
In total, each participant watches 28 videos, in a random order. 
After watching each video, the participant is asked to rate their experience on the scale of 1 to 5~\cite{MOS}.\footnote{\camera{We acknowledge that by showing a participant four versions of the same video, the participants may tend to scale their rates on the same video from the lowest to the highest. 
While we cannot entirely prevent it, we try to mitigate this potential bias by displaying the 28 videos in a random order, so the different versions of the same video are rarely displayed one after another, which reduces the chance that a participant scales his/her rating in a certain way.
 \questions{C.1}}}
\camera{For each video, we randomly pick a viewpoint trajectory from the 48 real traces and record the video {\em as if~} the user's viewpoint moves along the picked trajectory with the quality level picked by \name or the baseline.
That means Pano can still use its viewpoint prediction to adapt quality over time.
The participants watch these recorded videos on an Oculus headset~\cite{Oculus} (which generates real DoF and luminance changes).
They are advised not to move their viewpoints. 
Admittedly, this does not provide the exact same experience as the users freely moving their viewpoints. 
However, since each video is generated with real dynamic viewpoint trajectories, the experience of the users would be the same if they moved their viewpoints along the recorded trajectory. 
Additionally, this method ensures the participants rate the same videos and viewpoint trajectories across different streaming systems and bandwidth conditions. \questions{B.12}}

\mypara{Network throughput traces}
To emulate realistic network conditions, we use two throughput traces (with average throughput at 0.71Mbps and 1.05Mbps, respectively) collected from a public 4G/LTE operator~\cite{networktrace}. 
\camera{We pick these two throughput traces, because they are high enough to allow \name and the baselines to use high quality where users are sensitive (\eg areas with low JND), but not too high that all tiles can be streamed in the highest quality.
\questions{C.4}}

\begin{figure}
  \centering
  \begin{minipage}[t]{0.9\linewidth}
\centering
\includegraphics[width=\linewidth]{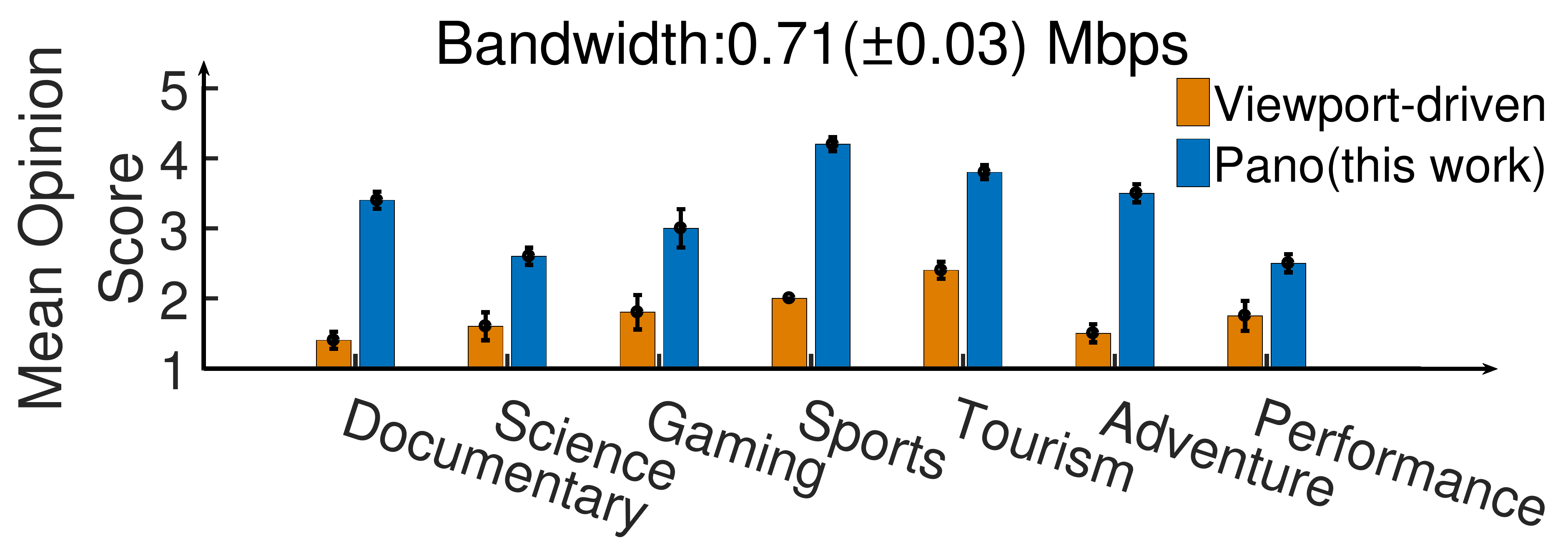}
\label{fig:side:a}
\end{minipage}
\begin{minipage}[t]{0.9\linewidth}
\centering
\includegraphics[width=\linewidth]{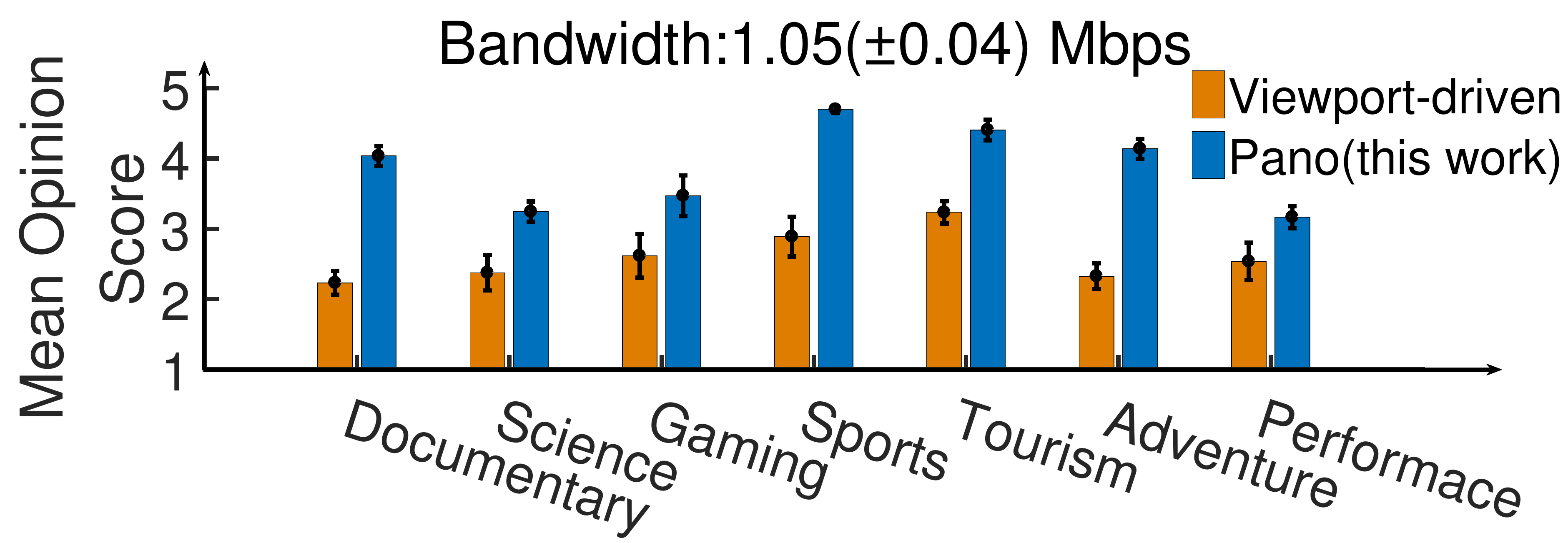}
\label{fig:side:b}
\end{minipage}
\vspace{-0.1cm}
  \tightcaption{Real user rating: \name vs. viewport-driven streaming. The figure summarizes the results of 20 users with the error bars showing the standard error of means.}
  \label{fig:rating}
\end{figure}

\begin{figure}
  \centering
    \begin{minipage}[t]{0.45\linewidth}
\centering
\includegraphics[width=0.6\linewidth]{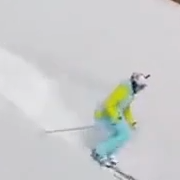}\\
{\small (a) Pano}
\label{fig:side:aa}
\end{minipage}\hspace{0.5cm}
   \begin{minipage}[t]{0.45\linewidth}
\centering
\includegraphics[width=0.6\linewidth]{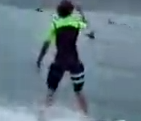}\\
{\small (b) Viewport-driven}
\label{fig:side:aa}
\end{minipage}
  \tightcaption{A snapshot of \vrvideo streamed by \name and Viewport-driven baseline.}
  \label{fig:show}
 \end{figure}

\begin{figure*}
  \centering    
  \begin{minipage}[t]{0.24\linewidth}
\centering
\includegraphics[width=\linewidth]{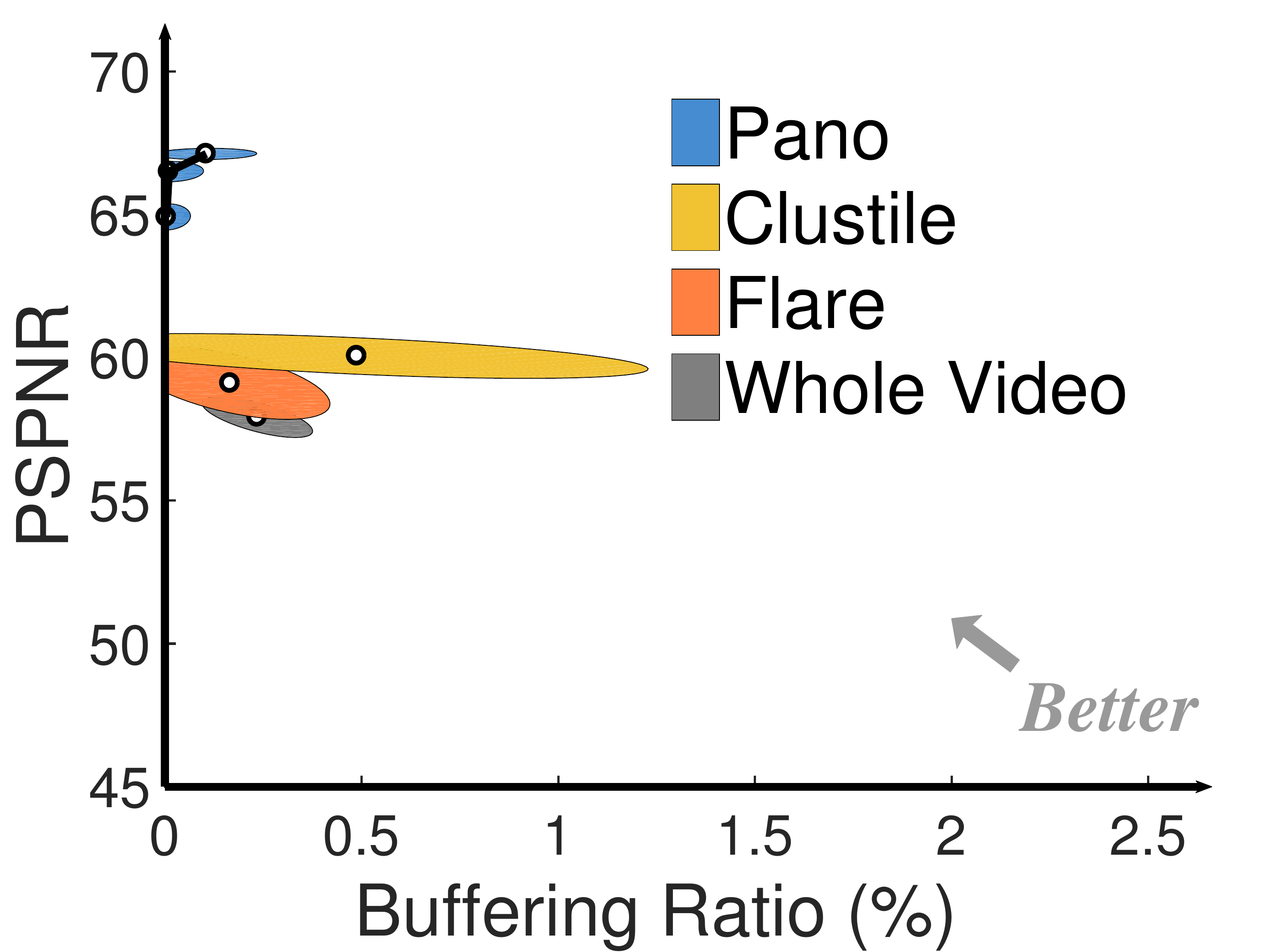}
{\small (a) Sports, Trace \#1}
\label{fig:side:a}
\end{minipage}\hspace{0.1cm}
  \begin{minipage}[t]{0.24\linewidth}
\centering
\includegraphics[width=\linewidth]{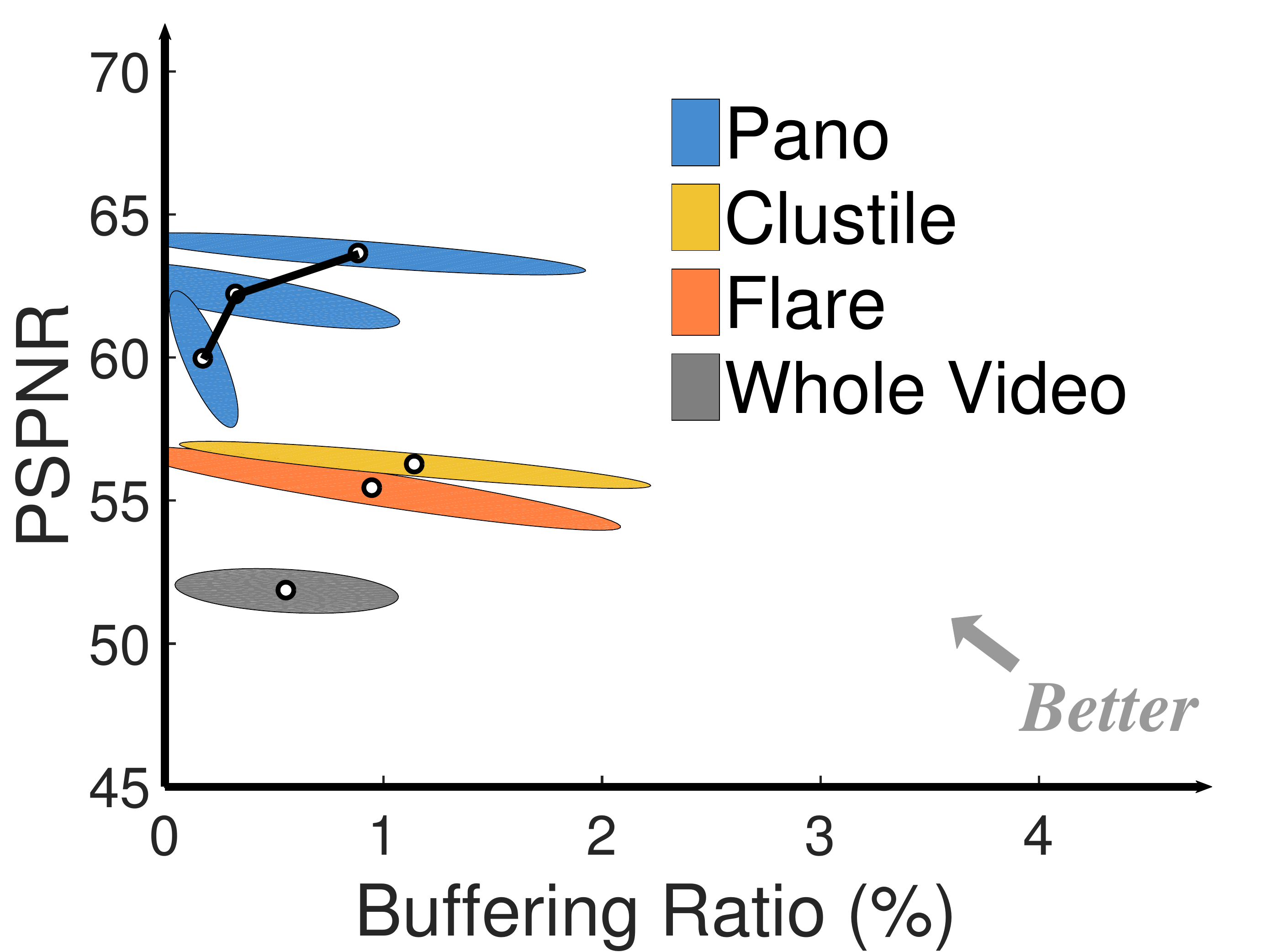}
{\small (b) Tourism, Trace \#1}
\label{fig:side:a}
\end{minipage}\hspace{0.1cm}
  \begin{minipage}[t]{0.24\linewidth}
\centering
\includegraphics[width=\linewidth]{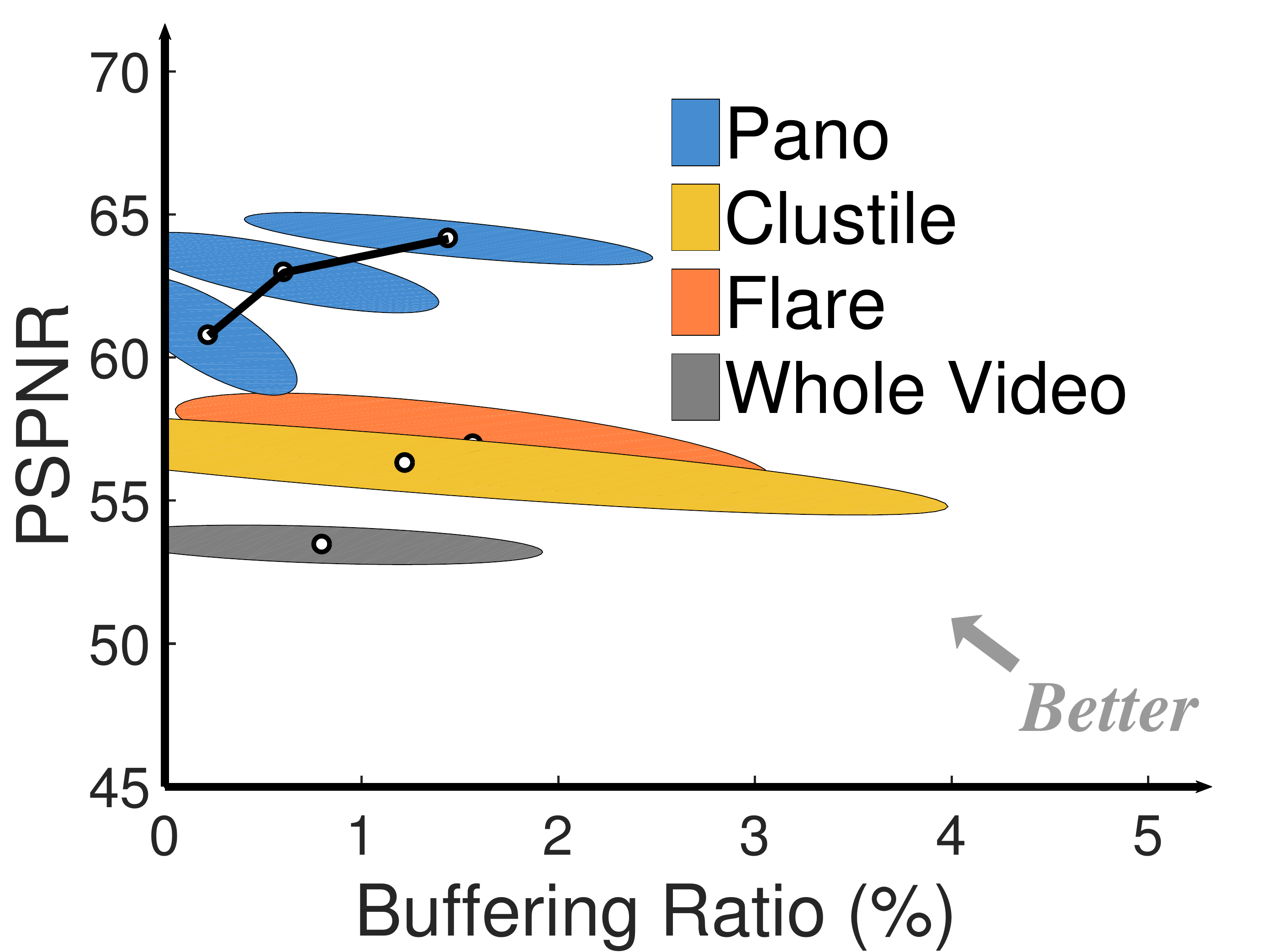}
{\small (c) Documentary, Trace \#1}
\label{fig:side:a}
\end{minipage}\hspace{0.1cm}
  \begin{minipage}[t]{0.24\linewidth}
\centering
\includegraphics[width=\linewidth]{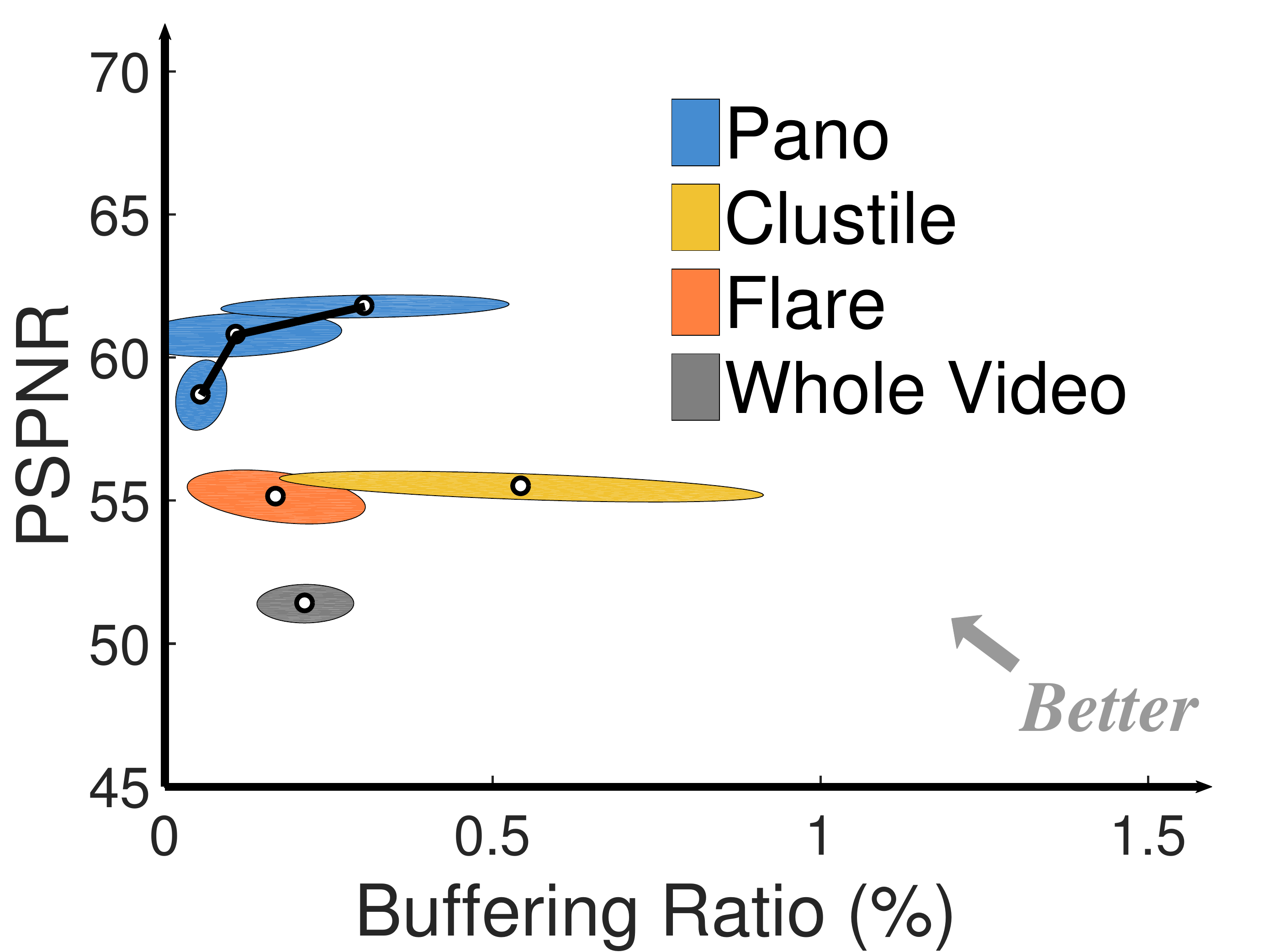}
{\small (d) Performance, Trace \#1}
\label{fig:side:a}
\end{minipage}\hspace{0.1cm}
   \begin{minipage}[t]{0.24\linewidth}
\centering
\includegraphics[width=\linewidth]{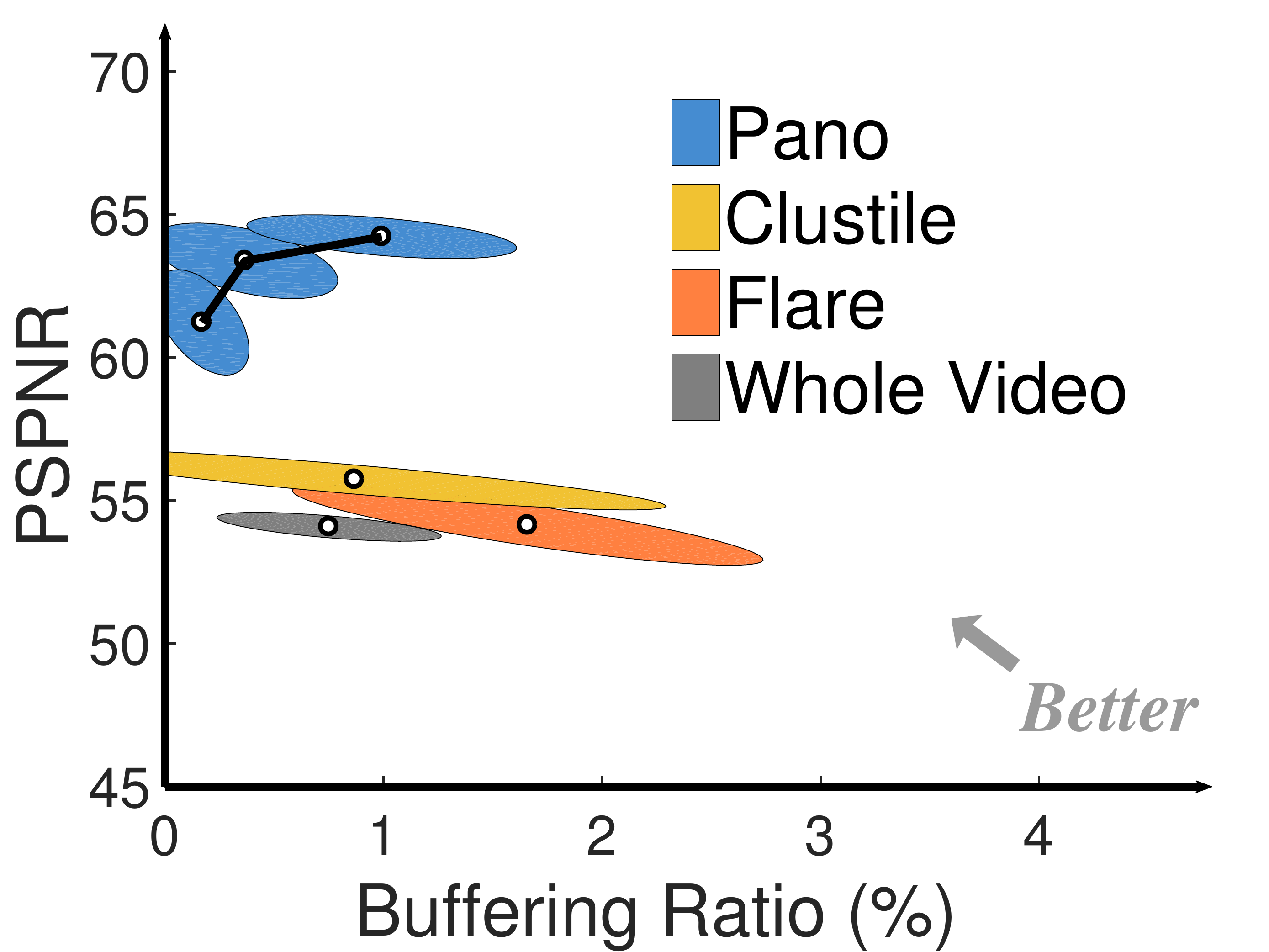}
{\small (e) Sports, Trace \#2}
\label{fig:side:a}
\end{minipage}\hspace{0.1cm}
  \begin{minipage}[t]{0.24\linewidth}
\centering
\includegraphics[width=\linewidth]{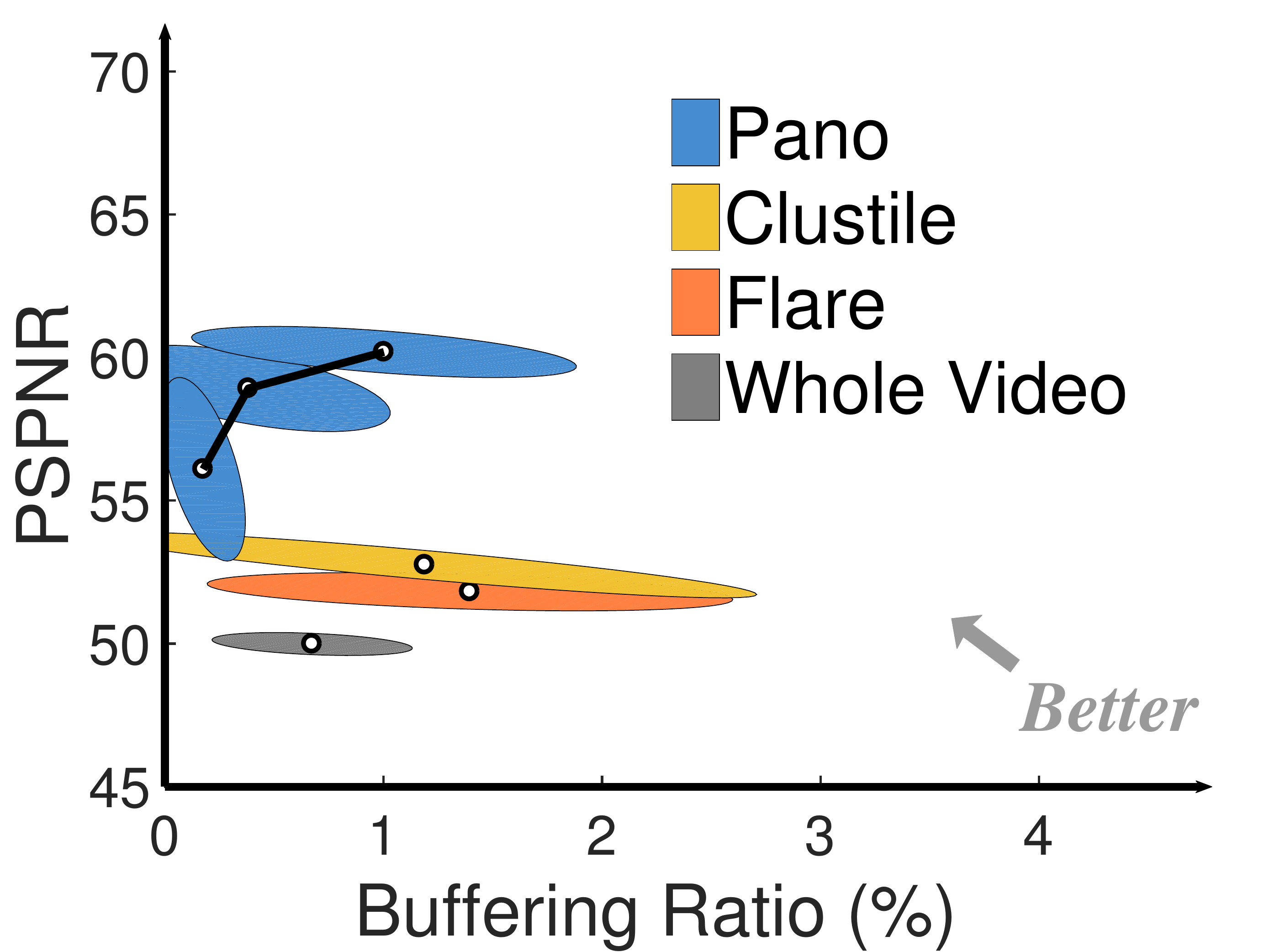}
{\small (f) Tourism, Trace \#2}
\label{fig:side:a}
\end{minipage}\hspace{0.1cm}
  \begin{minipage}[t]{0.24\linewidth}
\centering
\includegraphics[width=\linewidth]{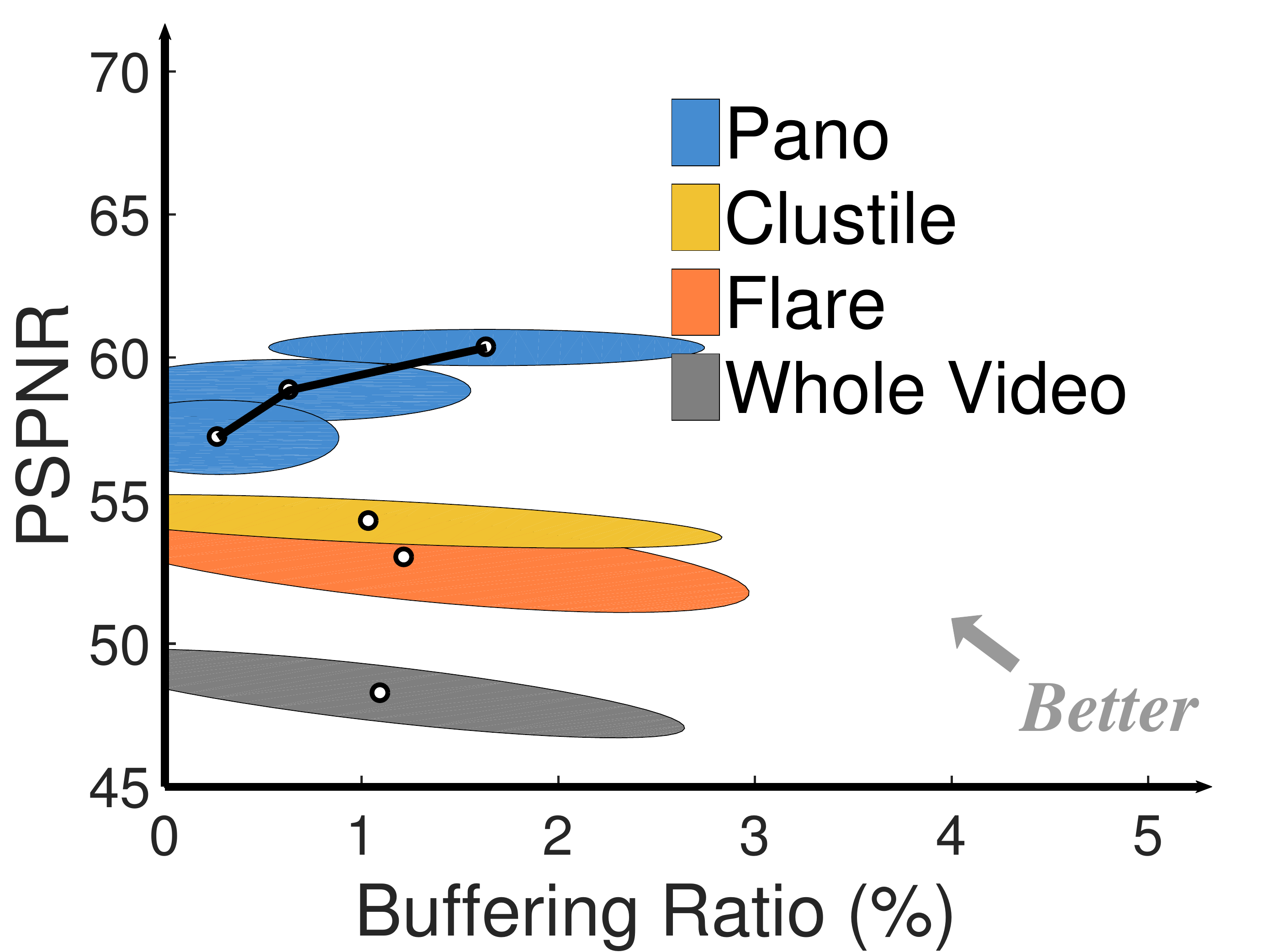}
{\small (g) Documentary, Trace \#2}
\label{fig:side:a}
\end{minipage}\hspace{0.1cm}
  \begin{minipage}[t]{0.24\linewidth}
\centering
\includegraphics[width=\linewidth]{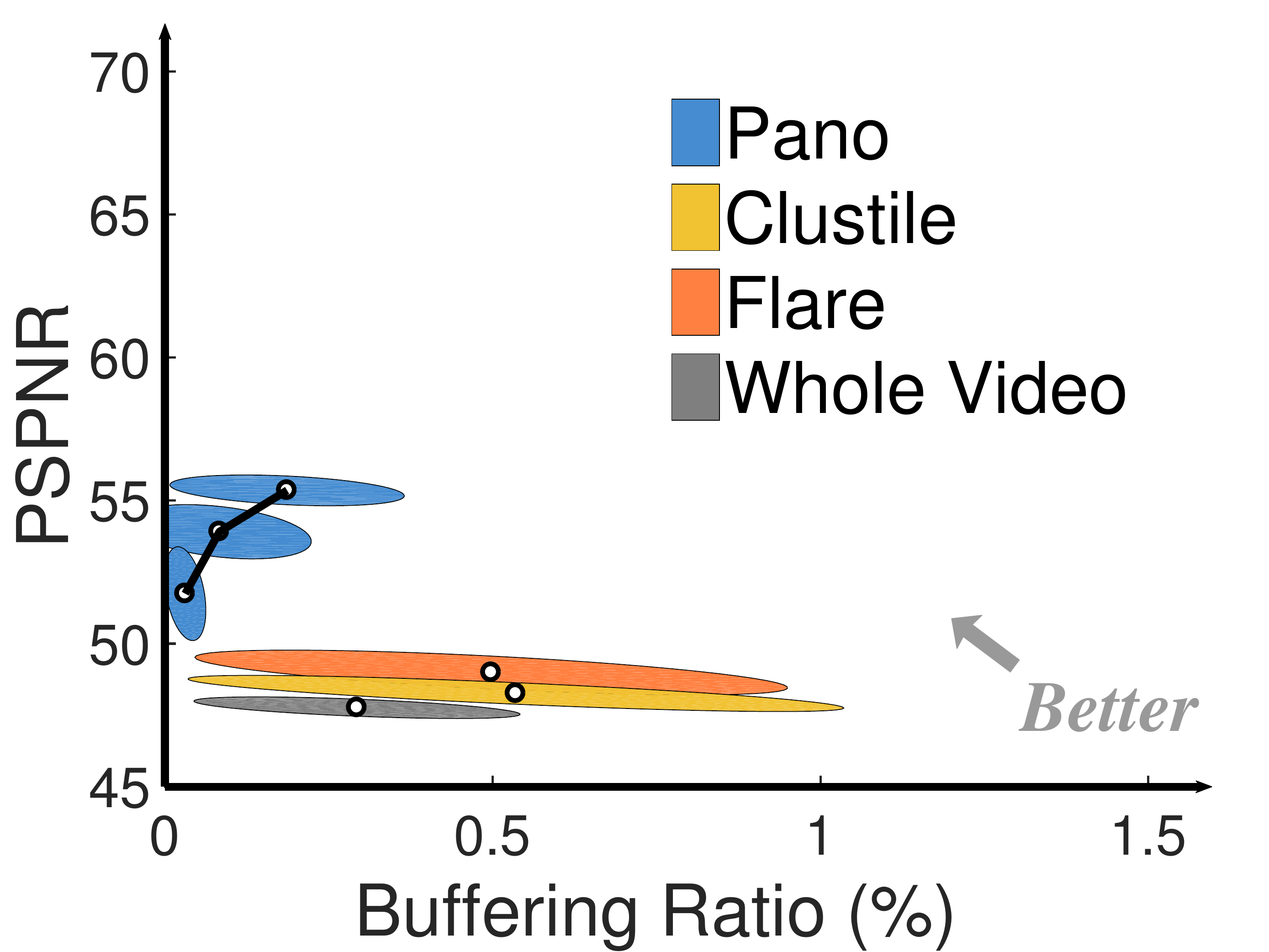}
{\small (h) Performance, Trace \#2}
\label{fig:side:a}
\end{minipage}\hspace{0.1cm}
  \tightcaption{Trace-driven simulation of four video genres over two emulated cellular links. Ellipses show 1-$\sigma$ range of results. We test \name with three target buffer lengths of \{1, 2, 3\} seconds~\cite{BOLA}.
  }
  \label{fig:overall}
\end{figure*}


\mypara{Quality metrics} We evaluate the video quality along two metrics that have been shown to be critical to user experience:
{\bf PSPNR}, and {\bf buffering ratio}.
We have seen PSPNR has a stronger correlation with \vrvideo user rating than alternative indices (Figure~\ref{fig:pspnr-accuracy}). 
Table~\ref{tab:pspnr-mos} maps the PSPNR ranges to corresponding MOS values. 
We define buffering ratio by the fraction of time the user's actual viewport is not completely downloaded.

\tightsubsection{End-to-end quality improvement}

\mypara{Survey-based evaluation}
Figure~\ref{fig:rating} compares the MOS of \name and the viewport-driven baseline (Flare) on the seven \vrvideos. 
\name and the baseline use almost the same amount of bandwidth (0.71Mbps or 1.05Mbps). 
We see that \name receives a much higher user rating, with 25-142\% improvement.
Figure~\ref{fig:show} shows the same snapshot under the two methods. 
The viewport-driven baseline gives equally low quality to both the moving object (skier) and the background of the viewport (not shown).
In contrast, \name detects the user is tracking the skier and assigns higher quality in and around the skier while giving lower quality to the static background (which appears to move quickly to the user).

\mypara{Trace-driven simulation}
Figure~\ref{fig:overall} compares \name with the three baselines on 18 videos across four content genres and over two network traces. 
Across all combinations, \name achieves higher PSPNR (user perceived quality), lower buffering ratio, or both.
We also see that \name has more improvement in some videos than others. 
This is due largely to the different levels of viewpoint dynamics across the videos. 
More dynamic viewpoint movements mean lowered sensitivities to quality distortion, thus more opportunities for \name to reduce quality levels without hurting users' perceived quality.
%

\tightsubsection{Robustness}



\begin{figure*}
\centering
\begin{minipage}[t]{0.235\textwidth}
\centering
\includegraphics[width=\linewidth]{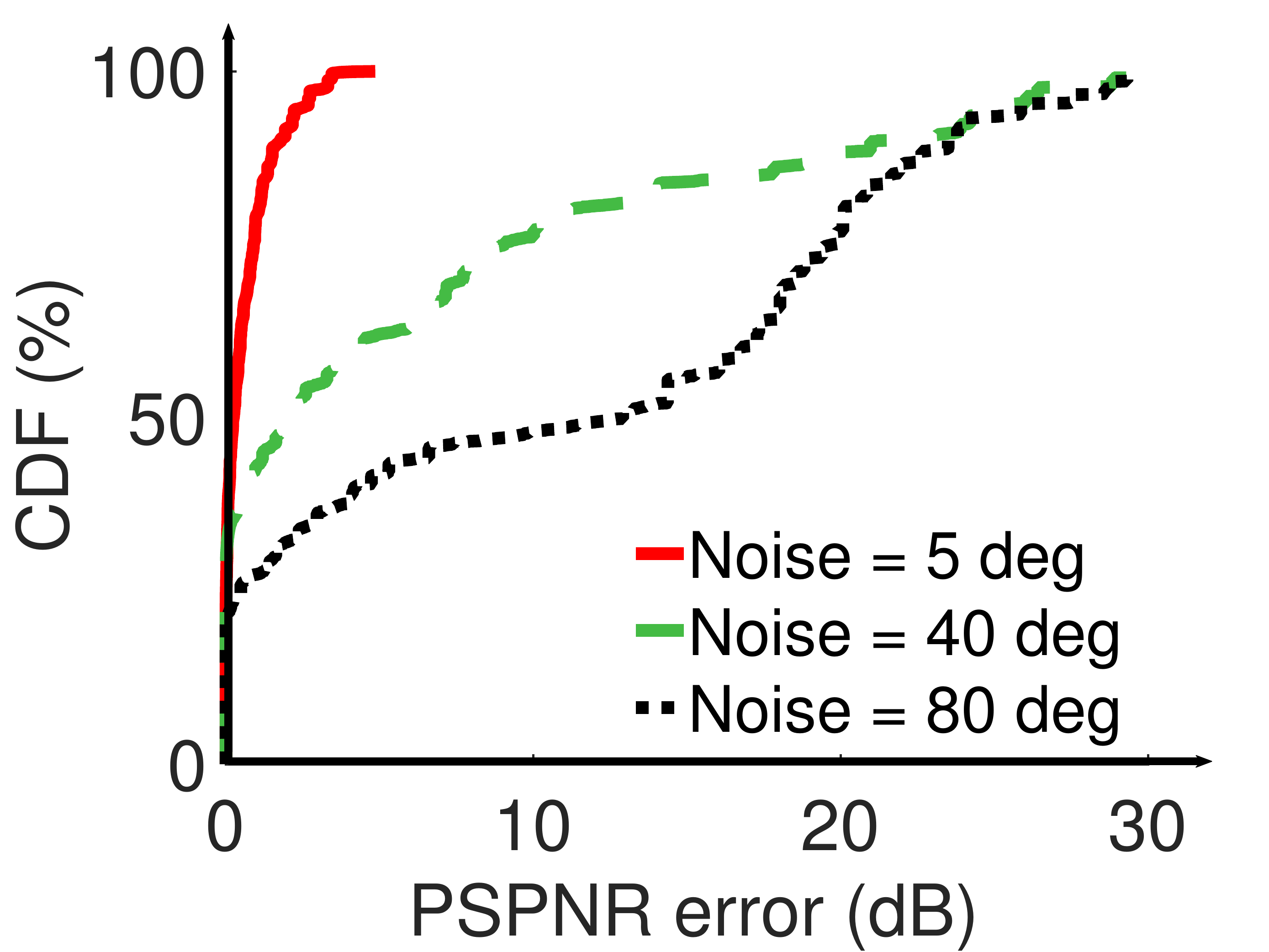}
{\small (a) PSPNR errors under noisy viewpoint prediction}
\label{fig:robust:error-distribution}
\end{minipage}
\begin{minipage}[t]{0.235\textwidth}
\centering
\includegraphics[width=\linewidth]{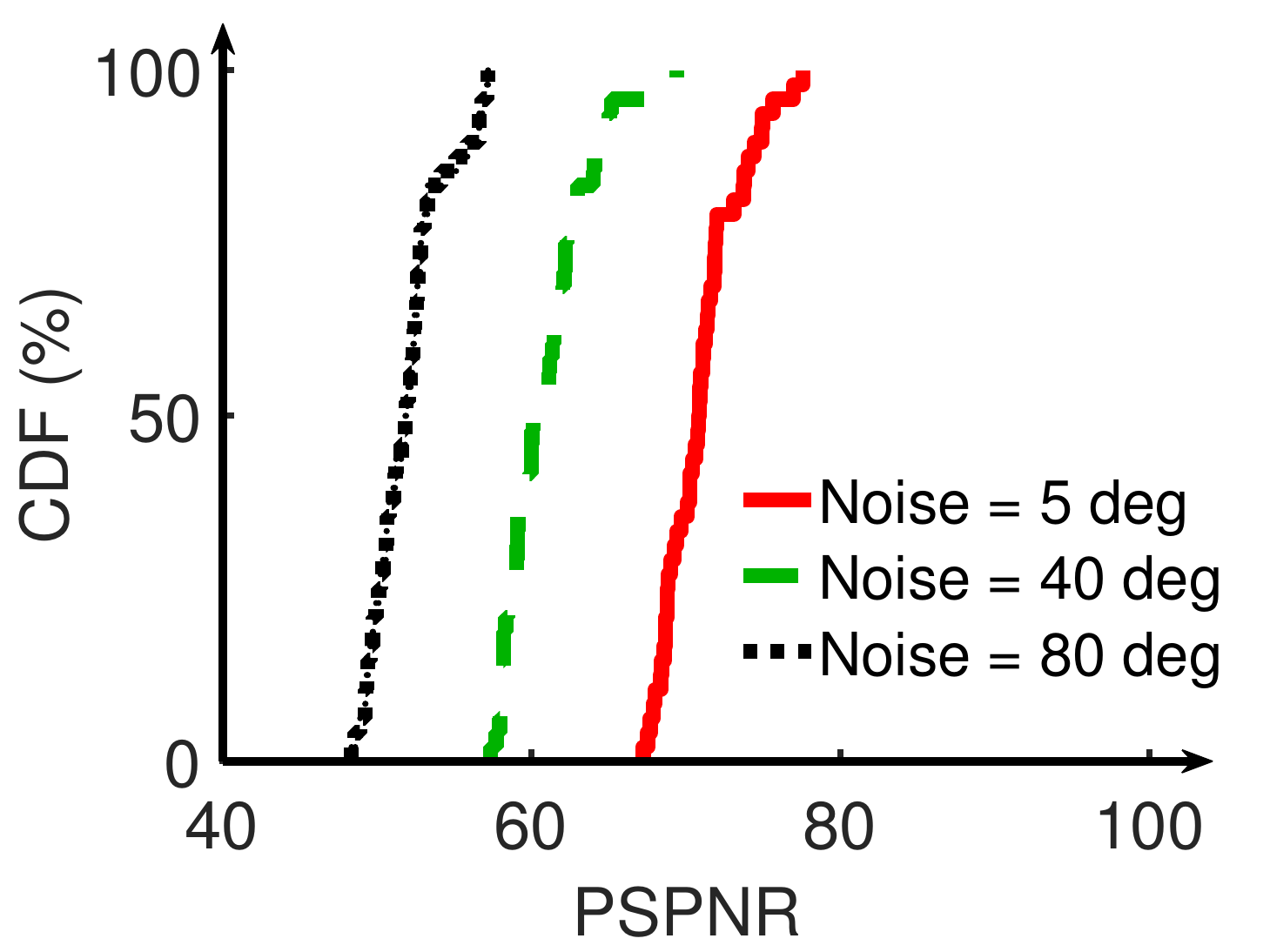}
{\small (b) Quality (PSPNR) distributions across users}
\label{fig:robust:quality-distribution}
\end{minipage}
\begin{minipage}[t]{0.235\textwidth}
\centering
\includegraphics[width=\linewidth]{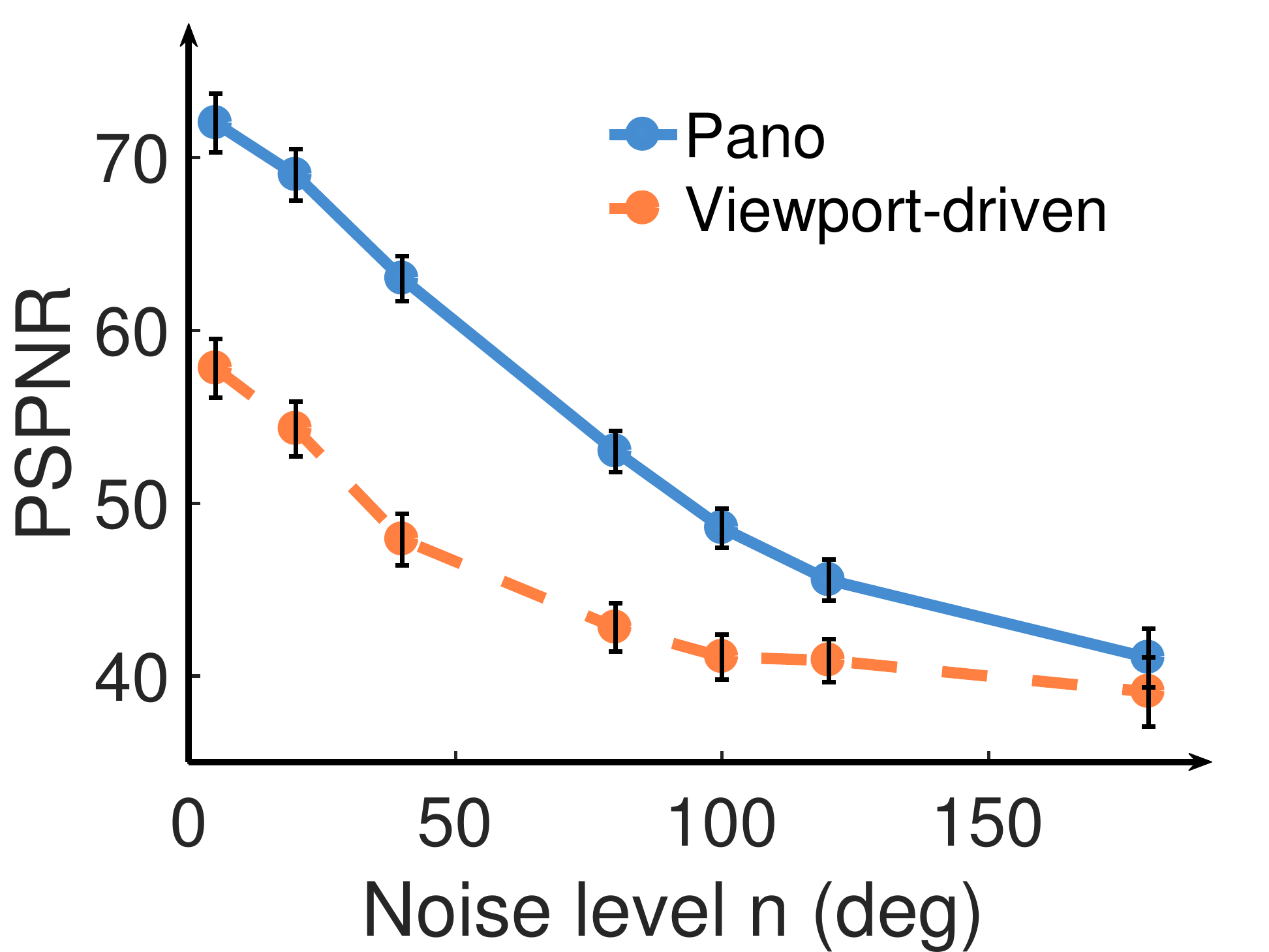}
{\small (c) Impact of viewpoint prediction errors on quality}
\label{fig:robust:quality-error}
\end{minipage}
\begin{minipage}[t]{0.235\textwidth}
\centering
\includegraphics[width=\linewidth]{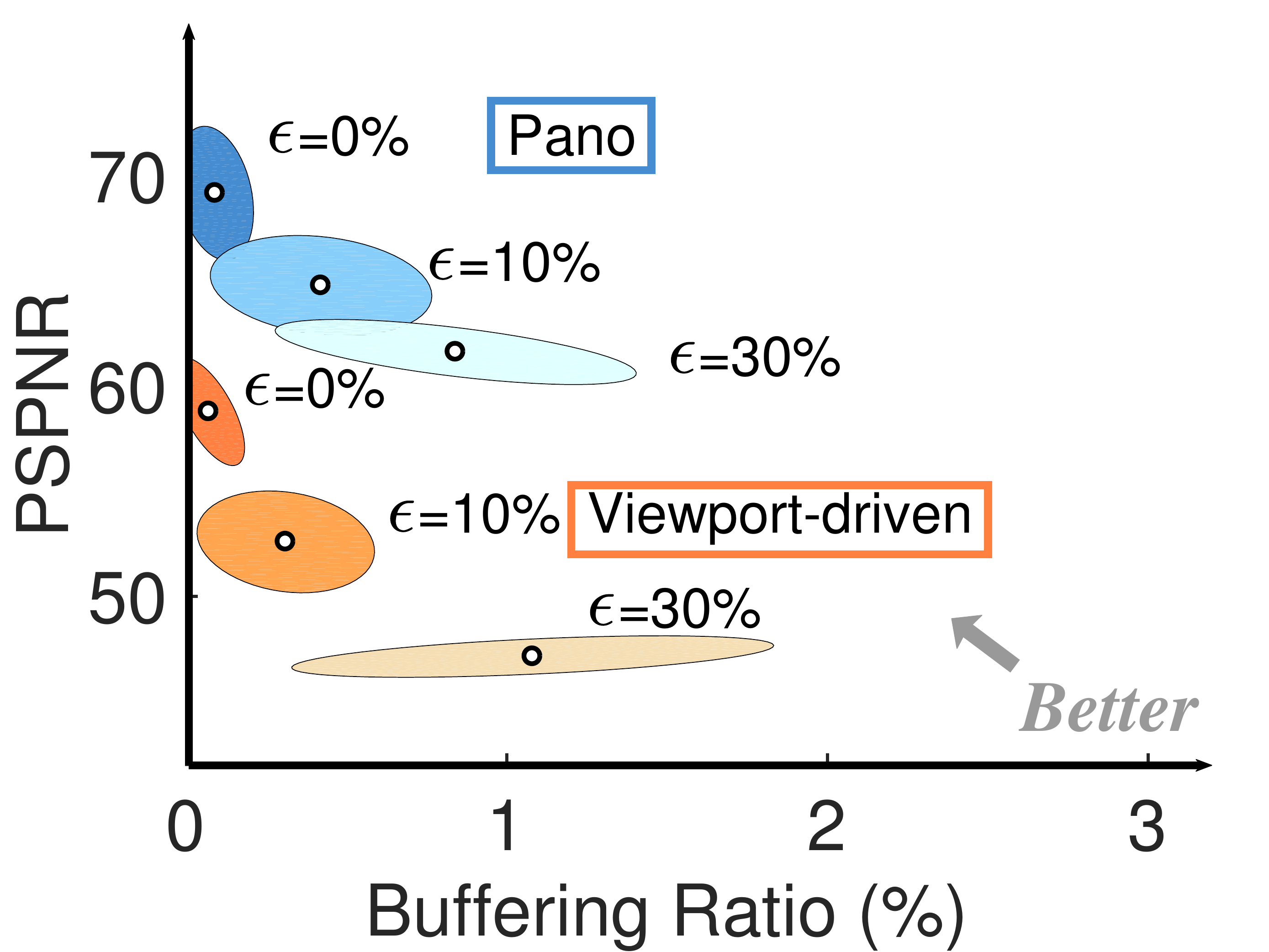}
{\small (d) Impact of bandwidth prediction errors on bandwidth-quality tradeoffs}
\label{fig:robust:bw-error-impact}
\end{minipage}
\vspace{-0.2cm}
\tightcaption{\camera{
\name is sensitive to noises of viewpoint movements. 
To stress test it, a random difference in degree is added to each actual viewpoint location, in order to increase viewpoint prediction errors. 
With higher viewpoint prediction errors, (a) \name estimates perceived quality (PSPNR) less accurately, and (b) the average perceived quality  drops (though with relatively small variance across users).
However, when compared to the viewport-driven baseline, (c) \name still achieves much higher perceived quality, though with diminishing gains as the viewpoint noise increases.
We also see that (d) \name is consistently better than the baseline under inaccurate bandwidth prediction.
\questions{A.3}\questions{A.7}\questions{D.1}}}
\label{fig:robust-1}
\end{figure*}

\mypara{Impact of viewpoint prediction noises}
To stress test \name under different viewpoint prediction errors, we create a noisier viewpoint trajectory from each real viewpoint trajectory in our trace, by adding a random shift to each actual viewpoint location.
Specifically, we shift the original viewpoint location by a distance drawn uniformly randomly between 0 and $n$ degrees, in a random direction. 
By increasing $n$, we effectively increase the viewpoint prediction errors.
Figure~\ref{fig:robust-1}(a) shows that more viewpoint noise ($n$) does reduce the PSPNR prediction accuracy, but the impact is not remarkable; a 40-degree noise only deviates the median PSPNR prediction by 7dB. 
This corroborates the intuition in \S\ref{subsec:adaptation} that \name's PSPNR prediction can tolerate a small amount of noise in viewpoint movement.
Moreover, Figure~\ref{fig:robust-1}(b) shows that the average perceived quality does drop with higher viewpoint prediction error, \camera{but the quality always has relatively small variance across users. 
This suggests that all users, including those whose viewpoint trajectories are very different from the majority, have similar perceived quality.\questions{A.3}}
\camera{Figure~\ref{fig:robust-1}(c) shows that \name consistently outperforms the baseline under an increasing level of viewpoint noise, although with diminishing improvements. \questions{D.1}}
\camera{Because subjective rating (MOS) is monotonically correlated with PSPNR (Table~\ref{tab:pspnr-mos}), we expect that \name's MOS would be similarly better than that of the baseline, despite the presence of viewpoint noises. \questions{D.1} 
\questions{A.3}\questions{A.7}}

%


\mypara{Impact of throughput prediction errors}
Figure~\ref{fig:robust-1}(d) shows the performance of \name (in PSPNR and buffering ratio) under different throughput prediction errors (a prediction error of 30\% means the predicted throughput is always 30\% higher or lower than the actual throughput).
We can see that as the throughput prediction error increases, \name's quality degrades, but the degradation is similar to that of the viewport-driven baseline (Flare). 
This is because \name consumes less bandwidth to provide the same perceived quality, which is robust when throughput drops down dramatically.

\tightsubsection{System overhead}
Next, we examine the overheads of a \vrvideo streaming system, in computing overhead, video start-up delay, and server-side preprocessing delay. 
We use an Oculus headset (Qualcomm Snapdragon 821 CPU, 3GB RAM, Adreno 530 GPU) as the client, a Windows Server 2016-OS desktop of Intel Xeon E5-2620v4 CPU, 32GB RAM, Quadro M2000 GPU as the video provider, and a 5-minute sports video as the test video.

\begin{figure}
\centering
 \begin{minipage}[t]{0.31\linewidth}
\centering
\includegraphics[width=\linewidth]{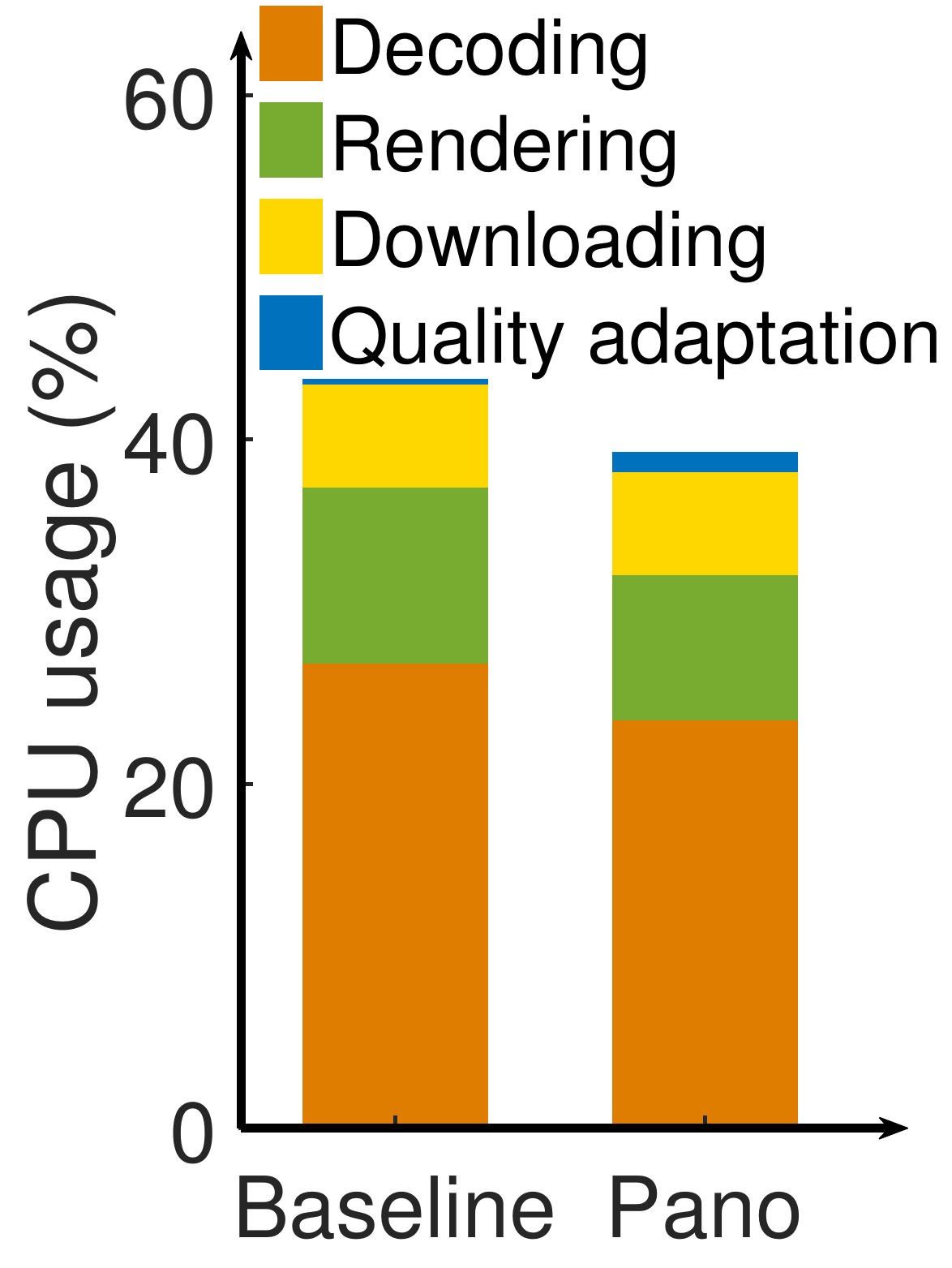}
{\small (a) Client-side CPU overhead}
\label{fig:side:a}
\end{minipage}
\begin{minipage}[t]{0.31\linewidth}
\centering
\includegraphics[width=\linewidth]{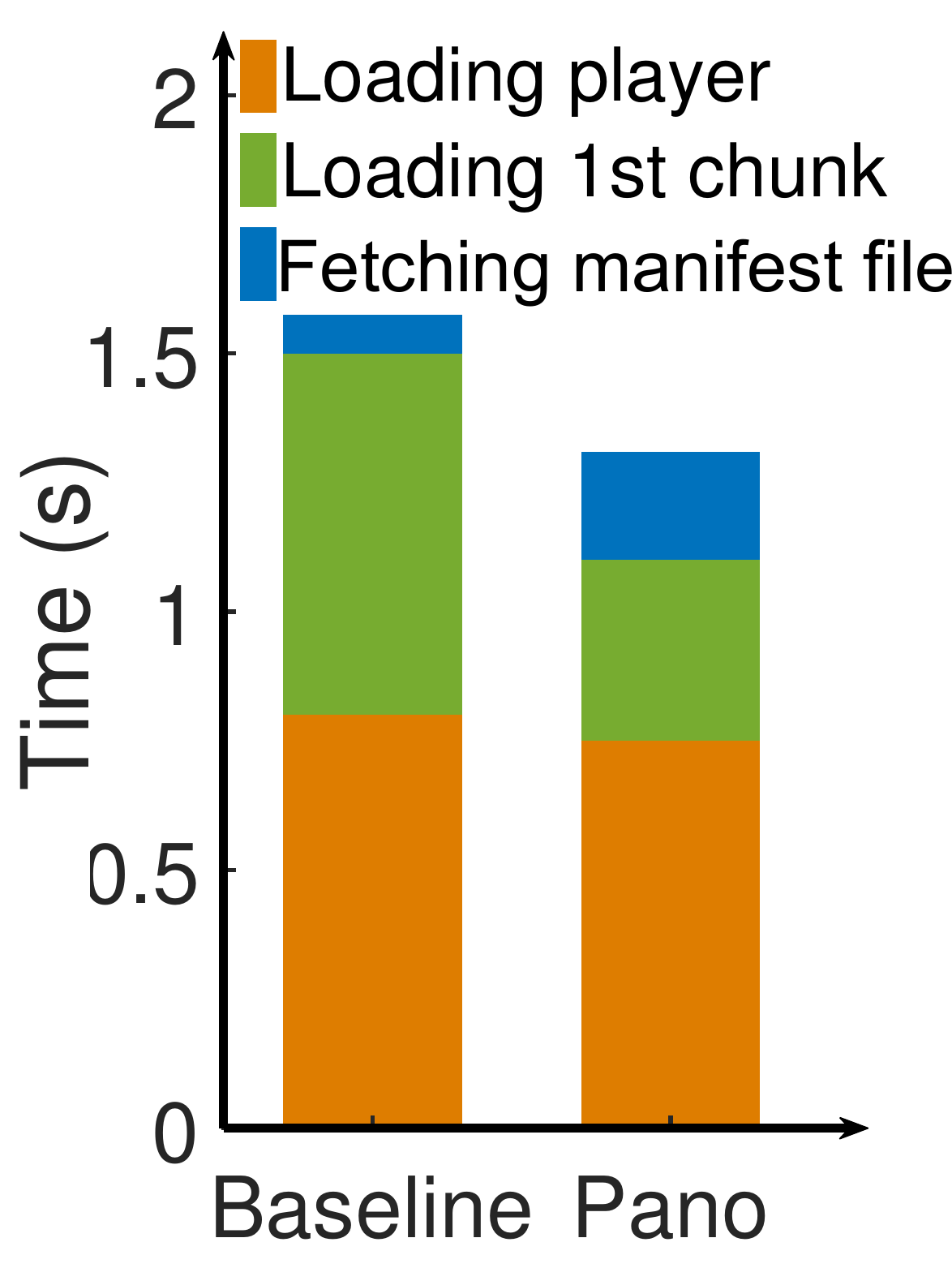}
{\small (b) Video start-up delay}
\label{fig:side:b}
\end{minipage}
\begin{minipage}[t]{0.31\linewidth}
\centering
\includegraphics[width=\linewidth]{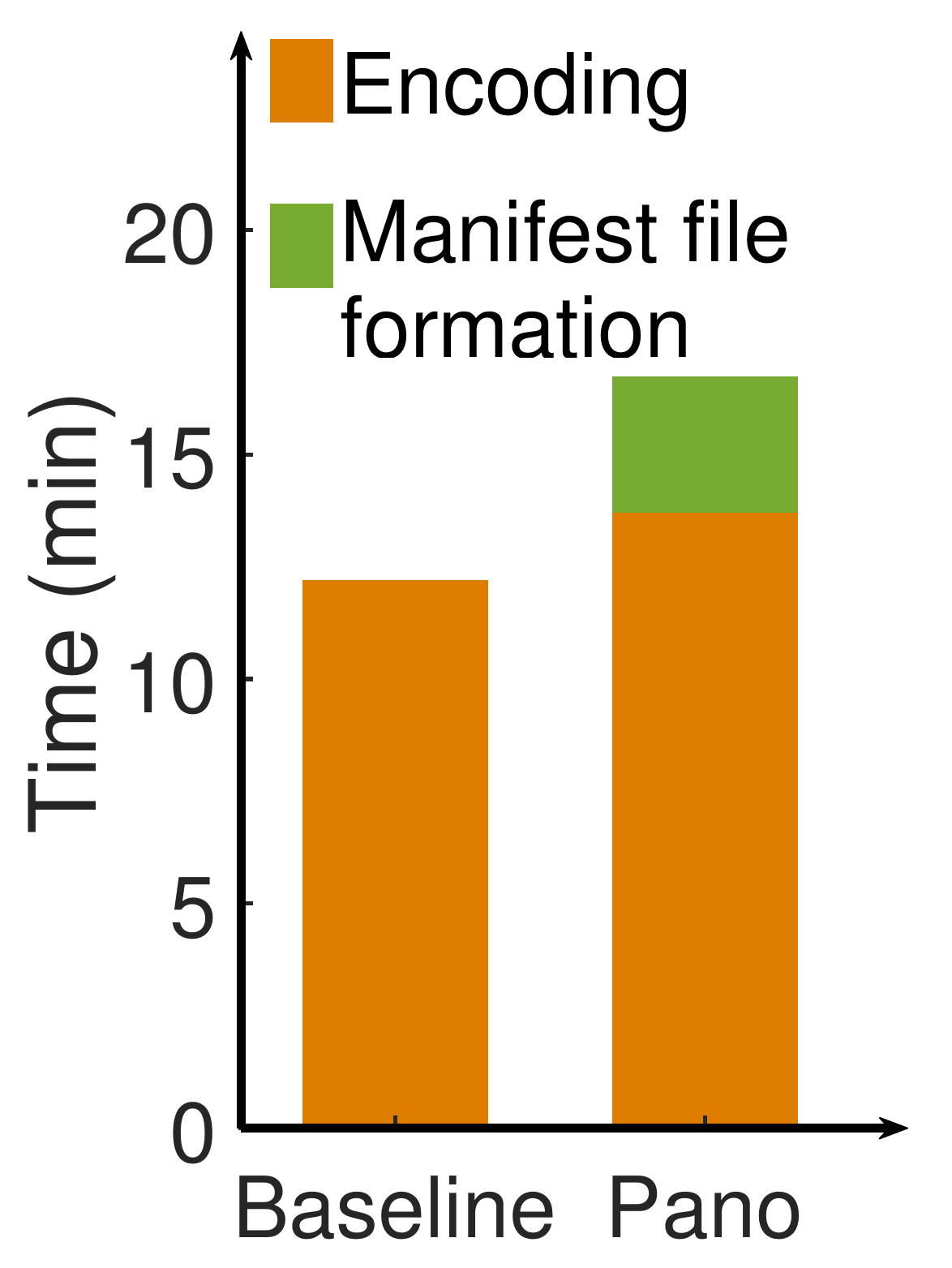}
{\small (c) Pre-processing time for a 1-minute video}
\label{fig:side:c}
\end{minipage}
\vspace{-0.1cm}
\tightcaption{\name reduces client-side processing overhead (a) and start-up delay (b) with minimal additional costs. The pre-processing time of \name is on par with the baseline (c). 
\vspace{-0.1cm}}
\label{fig:overhead}
\end{figure}

\mypara{Client-side overhead}
Figure~\ref{fig:overhead}(a) breaks down the client-side CPU overhead into that of four sequential steps: deciding per-tile quality level (quality adaptation), downloading, decoding, and rendering video tiles. 
We see that compared to the baseline of Flare, \name induces less computing overhead.
This is because \name needs to render video tiles with less total size than the baseline, and although \name needs extra PSPNR computation to make quality adaptation decisions, 
the client-side overhead is still dominated by video decoding and rendering, which is shared by both \name and the baselines. 

\mypara{Video start-up delay}
Figure~\ref{fig:overhead}(b) breaks down the video start-up delay (from when video player starts loading to when video starts playing) into three steps: loading the player, downloading the manifest file, and downloading the first chunk. 
Again, we see that \name induces an additional overhead since it needs to download a larger manifest file that includes the PSPNR lookup table (see \S\ref{sec:impl}).
However, the additional start-up delay is offset by the reduction of the loading time of the first chunk, because \name uses less bandwidth (to achieve the same PSPNR).

\mypara{Video processing overhead}
Figure~\ref{fig:overhead}(c) shows the pre-processing delay on the video provider side to pre-compute the PSPNR look-up table and encode the one minute worth of video (including chunking and tiling). Both the baseline and \name fully utilize the CPU cycles. 
\camera{Note that the preprocessing time does not include building the JND model. Because the \vrjnd model (as described in \S\ref{sec:jnd}) is agnostic to the specific video content, the \vrjnd model is generated once and used in all \vrvideos.
}We can see that \name does impose a longer pre-processing delay, due not only to the additional PSPNR pre-computation, but also to the variable-size tiling, which is more compute-intensive than the traditional grid-like tiling.
Nevertheless, the processing time of \name is still on par with the baseline.

\begin{figure}
\centering
 \begin{minipage}[t]{0.48\linewidth}
\centering
\includegraphics[width=\linewidth]{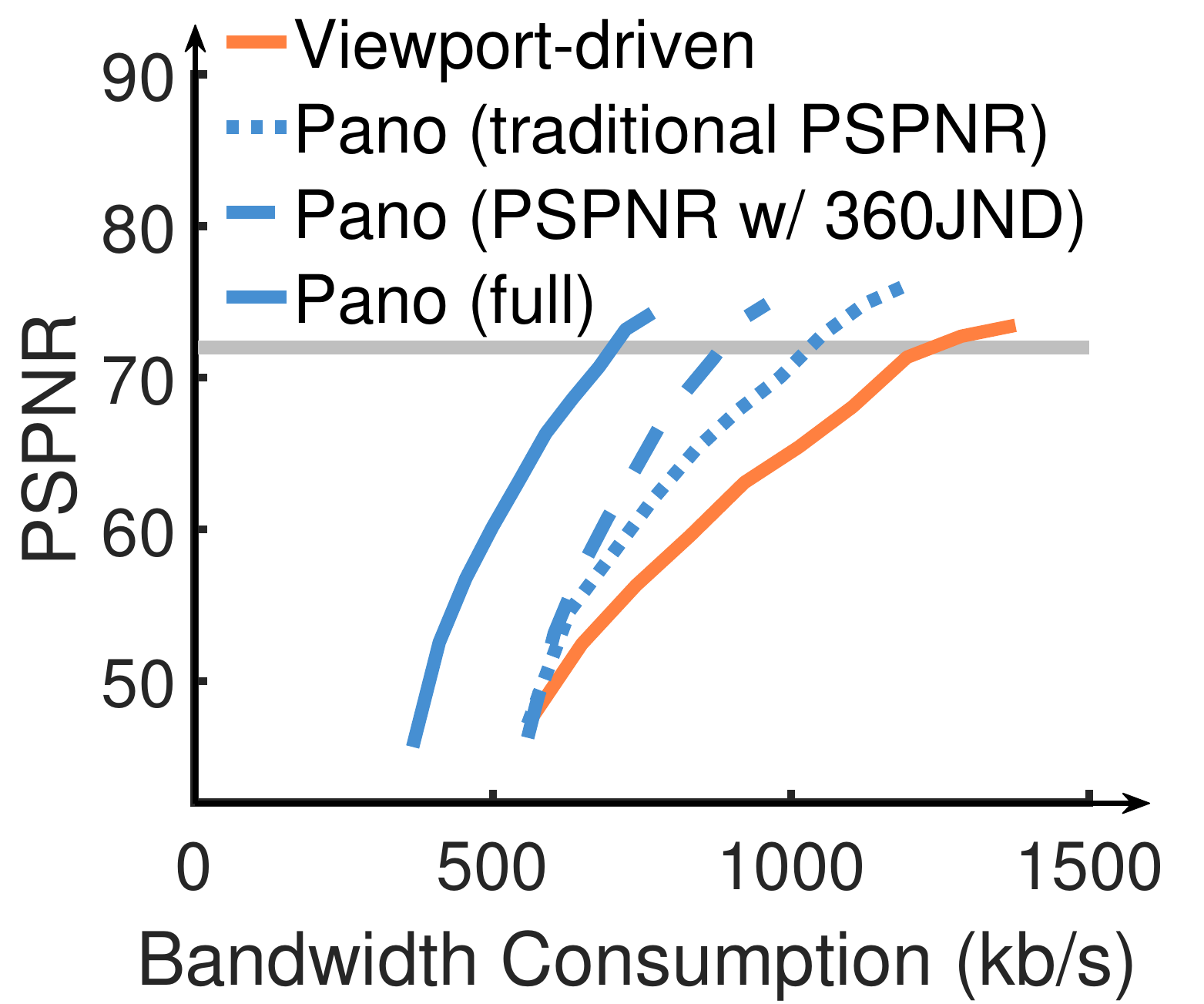}
{\small (a) Component-wise analysis.}
\label{fig:component}
\end{minipage}
\begin{minipage}[t]{0.48\linewidth}
\centering
\includegraphics[width=\linewidth]{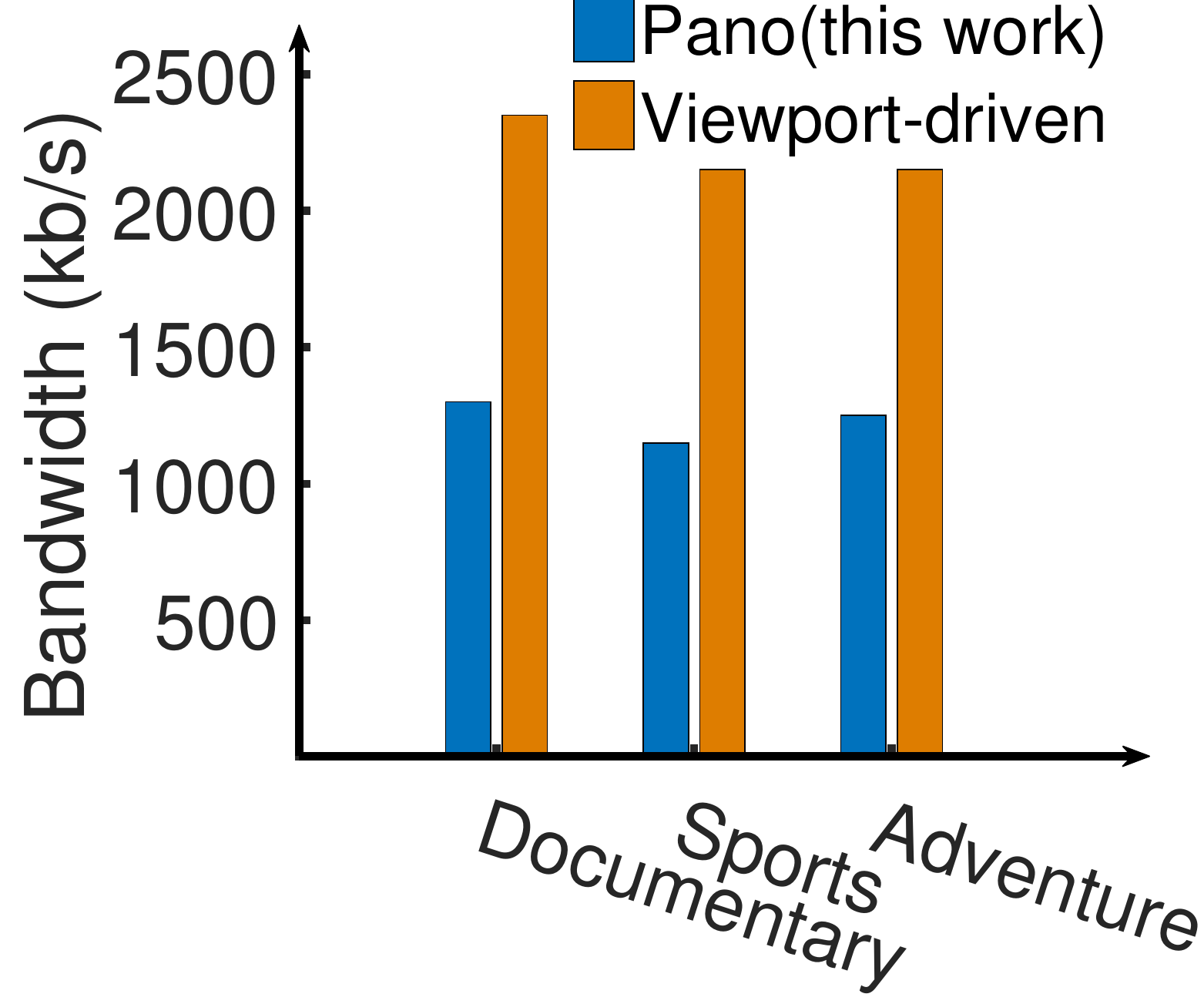}
{\small (b) Bandwidth consumption.}
\label{fig:bandwidthcomsumption}
\end{minipage}
\tightcaption{\name reduces the bandwidth consumption needed to achieve high quality (PSPNR = 72, or MOS = 5).
\vspace{-0.2cm}}
\label{fig:indepth}
\end{figure}

\tightsubsection{Bandwidth savings}
Finally, Figure~\ref{fig:indepth}(a) runs a component-wise analysis to evaluate the contribution of each technique in \name by adding one of them at a time to a viewport-driven baseline. 
To evaluate bandwidth savings on a larger set of videos, we extend our dataset from 18 \vrvideos to 50 \vrvideos (publicly available at~\cite{pano}), generate synthetic viewpoint traces for the new 32 \vrvideos as follows. 
We detect objects in each video using Yolo~\cite{yolov3}.
Then, we synthetically generate 48 viewpoint traces for each video by assuming that the viewpoint tracks a randomly picked object for 70\% of the time and looks at a randomly picked region for the remaining 30\% of the time.
\camera{We acknowledge that it may not be the most realistic way to model viewpoint trajectories, but we believe it is useful because (1) the bandwidth consumption is still derived from encoding real videos, and (2) the fraction of object-tracking time (70\%) matches the average object-tracking time in the real viewpoint traces.
\questions{B.10}}

Conceptually, we can breakdown the improvement of \name over the viewport-driven baseline (Flare) into three parts.
Figure~\ref{fig:indepth}(a) shows the bandwidth savings by each part, while holding the PSPNR to be 72 (which approximately translate to MOS = 5).
\camera{
\begin{packedenumerate}
\item {\bf Benefit of JND-awareness:} 
Switching from the basic viewport-driven quality model (\ie the perceived quality of a tile is only a function of its distance to the viewpoint) to a PSPNR-based quality model (which only includes the traditional JND-related factors~\cite{PSPNR,distance}) already saves 17\% of bandwidth. 
\item {\bf Benefit of \vrjnd vs. classic JND:} 
Next, if we add three new \vr-specific quality-determining factors into the PSPNR model (\S\ref{sec:jnd}) and quality adaptation (\S\ref{sec:control}), we can further save 11\% bandwidth consumption. 
\item {\bf Benefit of variable-size tiling:} 
Finally, the PSPNR-aware variable-size tiling (\S\ref{sec:tiling}) reduces the bandwidth consumption, over grid tiling, by another 17\%. 
\end{packedenumerate}
\questions{A.9}}

Finally, we run the evaluation with real throughput traces. 
Figure~\ref{fig:indepth}(b) shows that \name achieves the same PSPNR with 41-46\% less bandwidth consumption than the viewport-driven baseline. 

%

%% file: limitations.tex

\camera{\tightsection{Limitations of \vrjnd modeling}

 \questions{A.5}
Our \vrjnd model (\S\ref{sec:jnd}) is built on a survey study, where participants were asked to watch and rate their experience for videos that spanned a wide range of viewpoint speeds, DoF differences, and luminance changes. 
This is a similar methodology to what was used in the related work~\cite{PSPNR,distance}.
That said, we acknowledge two limitations of this approach. 

First, the values of \vrvideo-specific factors are varied in a specific manner (see details in Appendix), which may not match how they would vary and be perceived by users in the wild.
For instance, when we emulated different viewpoint moving speeds, the viewpoint was always moving in the horizontal direction and at a constant rate. 
However, when watching a \vrvideo, a user may move the viewpoint neither horizontally, nor at a constant speed.

Second, we have only tested the impact of two factors at non-zero values (Figure \ref{fig:two-factor}). 
We have not tested \vrjnd under all three factors at non-zero values.
Instead, we assume their effects on JND are mutually independent, thus could be directly multiplied (Equation~\ref{eq:pspnr}).
While Figure~\ref{fig:pspnr-accuracy} suggests our \vrjnd calculation is strongly correlated with user-perceived quality, \name could benefit from a more complete and fine-grained profiling of the relationship between \vrjnd and various factors.
\questions{C.5}
}

%% file: related.tex

\tightsection{Related work}
\label{sec:related}

\vrvideo streaming has attracted tremendous attention in industry~\cite{facebook,att, Google} and academia~\cite{Flare,POI360, Gaddam,Graf,Hosseini,Zare,CLS,ProbDASH,ban,Corbillon2017Viewport,Duanmu2017Prioritized, Sreedhar2017Viewport}. Here we survey the work most closely related to \name. 

\mypara{Viewport tracking}
Viewport-driven adaptation is one of the most popular approaches to  \vrvideos streaming~\cite{att,POI360,Gaddam,Zare,Corbillon2017Viewport, Duanmu2017Prioritized}. 
The viewport of a user is delivered with high quality, while other areas are encoded in low quality or not streamed.
To accommodate slight viewpoint movement, some work takes the recent viewport and re-scales it to a large region~\cite{POI360,Graf}, but it may still miss the real-time viewport if the viewport moves too much~\cite{Qian}.
To address this issue, many viewport-prediction schemes~\cite{Flare,CLS,ProbDASH,Afshin} are developed to extrapolate the user's viewport from history viewpoint movements~\cite{Qian}, cross-user similarity~\cite{ban}, or deep content analysis~\cite{Fan2017Fixation}. 
In addition to predict the viewpoint location, \name also predicts the new quality-determining factors (viewpoint-moving speed, luminance, and DoF) by borrowing ideas (\eg history-based prediction) from prior viewport-prediction algorithms. 

\mypara{\vrvideo tiling}
Tile-based \vrvideo encoding is critical for viewport-adaptive streaming~\cite{Gaddam,Zare,Corbillon2017Viewport,Flare,CLS}.
Panoramic video is spatially split into tiles, and each tile is encoded in multiple bitrates, so only a small number of tiles are needed to display the user's dynamic viewport. 
But this introduces additional encoding overhead as the number of tiles increases. 
Grid-like tiling is the most common scheme.
Alternative schemes, like ClusTile~\cite{ClusTile}, cluster some small tiles to one large tile so as to improve compression efficiency.
What is new in \name is that it splits the video in variable-size tiles which are well-aligned with the spatial distributions of the new quality-determining factors.

\mypara{Bitrate adaptation in \vrvideos}
Both \vrvideos and non-\vrvideos rely on bitrate-adaptation algorithms to cope with bandwidth fluctuations, but \vrvideos need to spatially allocate bitrate among the tiles of a chunk~\cite{ProbDASH,Flare} (tiles closer to the viewpoint get higher bitrates), but non-\vrvideos only change bitrate at the boundaries between consecutive chunks (\eg~\cite{MPC,BOLA,festive}).
While \name follows the tile-based bitrate adaptation, it is different in that the importance of each tile is dependent not only to its distance to the viewpoint, but users' sensitivities to its quality distortion.

\mypara{Just-Noticeable Distortion and perceived quality}
Many psychological visual studies (e.g.,~\cite{distance,PSPNR,brain}) have shown that the sensitivity of Human Visual System (HVS) can be measured by Just-Noticeable Distortion (JND)~\cite{Jayant}. 
JND has been used in other video quality metrics (\eg~\cite{PSPNR}) to quantify subjective user-perceived quality, but
most of the existing studies are designed for video coding and non-\vrvideos.
This work aims to leverage the impact of interactive user behaviors (such as viewpoint movements) on JND and how users perceive \vrvideo quality, to achieve higher \vrvideo quality with less bandwidth consumption.

%% file: appendix.tex
\camera{

\vspace{0.1cm}
\noindent Appendices are supporting material that has not been peer reviewed.

\section{Appendix}
\questions{A.5}\questions{D.2}

This section presents the detailed methodology of modeling \vrvideo JND.

\begin{table}[t]
\begin{small}
\begin{tabular}{rl}
\hline
{\bf Equipment}       & Oculus GO              \\ \hline
{\bf CPU}    & Qualcomm Snapdragon 821\\ \hline
{\bf Memory}       & 3GB\\ \hline
{\bf Screen Resolution} & 2560 $\times$ 1440\\ \hline
{\bf Refresh Rate} & 72Hz\\ \hline
{\bf Fixed pupil distance} & 63.5mm\\ \hline
\end{tabular}
\end{small}
\vspace{0.3cm}
\tightcaption{Headset parameters used in JND modeling
}
\vspace{-0.4cm}
\label{tab:hmd}
\end{table}


\subsection{Survey process}

The user study was based on 20 participants (age between 20 and 26). 
The same 20 participants also did the survey-based performance evaluation (\S\ref{sec:eval}), so the results could be affected by the limited size of the participant pool.\questions{C.2}
In all tests, the participants watch (synthetically generated or real) \vrvideos using an Oculus headset~\cite{Oculus}, of which the parameters are summarized in Table \ref{tab:hmd}.

Each participant was asked to watch a video with an increasing level of quality distortion (see the next section for how the quality distortion was added to a video). 
Every time the quality distortion increased, the participant was asked whether he or she could perceive the quality distortion. 
We define JND of a video by the average level of quality distortion that was perceivable for the first time, across the 20 participants.
We repeated this test with 43 artificially generated videos, and each participant watched the videos in a random order (which helped mitigate biases due to any specific playing order).



\subsection{Test videos}

Next, we explain 
(1) how we artificially generate videos with a controlled noise level added to an visual object, to emulate the effect of a specific level of quality distortion, and 
(2) how we emulate the viewpoint behavior such that the visual object would appear in the video with a specific relative moving speed, DoF difference, or background luminance change.

All test videos were generated by manipulating a basic video where a small square-shaped foreground object (64$\times$64 pixels) was located in the center of the screen. The object has a constant grey level of 50.
We refer to the foreground object by $U$.

\mypara{Adding controlled quality distortion}
To add a controlled quality distortion on $U$, we borrow a similar methodology from prior user study on JND~\cite{PSPNR,distance}.
We randomly picked 50\% pixels of $U$, and added a value of $\Delta$ to their values (grey level).
We made sure that the resulting pixel values were still within the range of 0 to 255. 
By varying the value of $\Delta$ from 1 to 205, we created a video with an increasing level of quality distortion on the foreground object $U$. 
The video was played to each participant until the distortion was perceived for the first time.


\mypara{Emulating the effect of relative viewpoint-moving speed}
To emulate the perception of quality distortion under a specific relative viewpoint-moving speed, we fixed a red spot at the center of the screen, and moved the foreground object $U$ horizontally at a specific speed of $v$. That is, $U$ and the red spot (viewpoint) has a relative moving speed of $v$.
The participant was asked to look at the red spot, and report whether he or she could perceive the quality distortion added onto the object $U$.
This process emulated the effect of viewpoint moving at a relative speed of $v$ to where the quality distortion occurred.
We tested viewpoint speeds from 0 deg/s to 20 deg/s.

\mypara{Emulating the effect of luminance changes}
To emulate the perception of quality distortion under certain luminance changes, each video began with the background luminance set to $g+l$, and then reduced to $g$ after 5 seconds. 
Right after the luminance was reduced to $g$, the object $U$ was shown with a gradually increasing amount of quality distortion. 
The participant was then asked to report as soon as the quality distortion was perceived. 
Although the report quality distortion may not be the true minimally perceivable quality distortion (JND), we found the participants always reported quality distortion within 3 seconds after luminance was reduced.
That suggests the first perceivable quality distortion might be a reasonable indicator of the real JND under the luminance change of $l$.
By fixing $g$ at 0 grey level (darkest) and varying $l$ from 0 to 240 grey level, we can test the JND under the different levels of luminance changes within a short time window of 5 seconds.

\mypara{Emulating the effect of DoF differences}
To emulate the perception of quality distortion on an object with a specific DoF difference from the viewpoint, we asked the participants to focus on a static spot displayed at a DoF difference $d$ ($d = \{0, 0.67, 1.33, 2\}$dioptre) from the foreground object $U$. 
Then quality distortion was added to the object $U$, and the participants were asked to report when they first perceived the quality distortion.

\mypara{Joint impact of two factors}
So far, each factor (relative viewpoint-moving speed, luminance change, DoF difference) was varied separately with others held to zero. 
We also tested the JND under both viewpoint speed and DoF differences at non-zero values simultaneously.
That is, at each possible relative viewpoint-moving speed, we enumerated different values of DoF differences, using the same method described above.
Similarly, we also tested the JND under both object luminance and relative viewpoint-moving speed at non-zero values. 
These results were shown in Figure~\ref{fig:two-factor}.

}